\documentclass[prb,aps,twocolumn,eqsecnum,amsmath,amssymb,showpacs]{revtex4}
\usepackage{graphicx}
\usepackage{dcolumn}
\usepackage{bm}

\begin{document}

\title{       Defect states and spin-orbital physics in doped vanadates:
              Y$_{1-x}$Ca$_x$VO$_3$}

\author {     Peter Horsch }
\affiliation{ Max-Planck-Institut f\"ur Festk\"orperforschung,
              Heisenbergstrasse 1, D-70569 Stuttgart, Germany }

\author {     Andrzej M. Ole\'{s} }
\affiliation{ Max-Planck-Institut f\"ur Festk\"orperforschung,
              Heisenbergstrasse 1, D-70569 Stuttgart, Germany, }
\affiliation{ Marian Smoluchowski Institute of Physics, Jagellonian
              University, Reymonta 4, PL-30059 Krak\'ow, Poland }

\date{8 July 2011}

\begin{abstract}
We present a model for typical charged defects in weakly doped
Y$_{1-x}$Ca$_x$VO$_3$ perovskites and study how they influence the
magnetic and orbital order. Starting from a multiband Hubbard
model we show that the charge carriers introduced by doping are
bound to the Ca defects with large binding energy  of $\approx 1$
eV at small doping, and give rise to the  in-gap absorption band
observed in the optical spectroscopy. The central position of a
generic Ca defect with eight equidistant vanadium neighbors
implies a partly filled defect band and permits activated
transport due to Coulomb disorder. We explore the effect of bound
charge carriers on the dynamics of the $\{yz,zx\}$ orbital and
spin degrees of freedom in the context of a microscopic
strong-coupling model. After deriving the superexchange interactions
around the doped hole we show that the transition from $G$-type
to $C$-type antiferromagnetic order is triggered by the kinetic
energy of doped holes via the double exchange mechanism. The defect
states lead to local modification of orbital correlations within
ferromagnetic chains along the $c$ axis --- some
of them contain hole defects while the charge-orbital coupling
suppresses locally $\{yz,zx\}$ orbital fluctuations in the others.
Thereby Ca defects provide a physical mechanism for spin-orbital
dimerization along the ferromagnetic bonds, suggesting that in the
$C$-AF phase of weakly doped Y$_{1-x}$Ca$_x$VO$_3$ dimerization
increases with doping.

\end{abstract}

\pacs{75.10.Jm, 71.10.Fd, 71.55.-i, 75.25.Dk}

\maketitle

\section{ Experimental motivation }
\label{sec:exp}

Recent experimental and theoretical investigations of transition
metal oxides have revealed the interplay between spin, charge and
orbital degrees of freedom, leading to dramatic changes of
magnetic and transport properties.\cite{Ima98} The best known
examples of this joint effect of several degrees of freedom are
high-temperature superconductivity and colossal magnetoresistance
(CMR). When orbital degrees of freedom contribute, as in the CMR
manganites, they may strongly influence magnetic properties 
and also may play a key role for the charge transport. \cite{Dag01,Feh04,Tok06}
Particularly in the manganites the strong coupling between 
orbital states and local lattice distortions plays a prominent
role, and orbital order is stabilized to a large extent by the
Jahn-Teller interactions.\cite{Fei99} Under these circumstances
quantum orbital fluctuations are suppressed and classical orbital 
order determines the spin structure and dynamics.\cite{Ole05} This
is most clearly displayed by the large difference between the temperatures 
of the orbital and magnetic phase transitions in LaMnO$_3$, 
being $T_{\rm OO}=780$ K and $T_{\rm N}=140$ K.\cite{Mur98} 

A challenge for the theory of spin-orbital systems and a qualitatively 
different situation is encountered in the $R$VO$_3$ perovskites, where 
$R$=Lu,Yb,$\cdots$,La. In these perovskites, i.e., controlled by 
$t_{2g}$ valence electrons, the orbital-lattice coupling is weak while 
the spin-orbital coupling is the dominant interaction and thus orbital 
fluctuations are not suppressed. A common feature is the onset of the 
$G$-type alternating orbital ($G$-AO) order below the characteristic 
orbital ordering temperature $T_{\rm OO}$ which is in these compounds 
relatively low, $T_{\rm OO}\simeq 200$ K, and comes close to the N\'eel 
temperature $100<T_{\rm N1}<140$ K for the magnetic transition to the
antiferromagnetic (AF) phase with AF order in $ab$ planes
accompanied by ferromagnetic (FM) order along the $c$ axis, the
$C$-AF phase.\cite{Miy03} Unlike $e_g$ orbitals in the CMR
manganites, in this class of compounds the $t_{2g}$ orbitals may 
form orbital ordered states which are subject to strong orbital quantum
fluctuations. The consequences of quantum spin-orbital interplay
in the $R$VO$_3$ perovskites were discussed in the theory,
\cite{Kha01,Hor03,Sir03,Kha04,Ole07,Hor08} and have been observed in 
several experiments.\cite{Miy02,Ulr03,Ree06,Goo07,Maz08,Zho09} The 
changes of spin and orbital correlations are responsible for the 
temperature dependence of the optical spectral weights,\cite{Kha04,Miy02} 
as well as for the dimerization of FM interactions in the $C$-AF phase 
observed in the neutron scattering in YVO$_3$,\cite{Ulr03} and also in 
LaVO$_3$.\cite{Tun08} Related instability of FM chains toward 
dimerization occurs at finite temperature when spin and orbital degrees 
of freedom couple.\cite{Sir08} Orbital fluctuations and their competion 
with orbital-lattice coupling play also a crucial role for the explanation 
of the nonmonotonous dependence of the orbital transition temperature 
on the radius $r_R$ of $R$ ions along the $R$VO$_3$ series.\cite{Hor08}

The phase diagram of the perovskite-type $R$VO$_3$ compounds\cite{Miy03}
shows several spin- and/or orbital ordered phases. In the regime
of compounds with low values of ionic radii $r_R$ of rare earth
ions $R$, another AF phase with complementary $G$-type AF ($G$-AF)
order\cite{Kaw94} accompanied by $C$-type alternating orbital ($C$-AO)
order (with staggered orbitals in $ab$ planes and repeated orbitals
along the $c$ axis) appears below the second magnetic transition at
$T_{\rm N2}$,\cite{Miy06} for example in YVO$_3$ $T_{\rm N2}=77$ K.
\cite{Ren00,Nog00,Bla01} In addition, recent Raman
experiments\cite{Miy06} suggest that the short-range orbital
fluctuations of the $G$-type occur in this intermediate $C$-AF phase
--- they coexist with the $C$-AO order and make it thus quite
different from the one observed in LaVO$_3$. The magnetic exchange
constants which determine the magnons in the $C$-AF phase  are
about a factor two smaller than those found in the low-temperature
$G$-AF phase.\cite{Ulr03} Therefore, it has been argued that this
phase transition in YVO$_3$ follows from large entropy of spin and
orbital excitations in the $C$-AF phase,\cite{Kha01,Hor03} but the
observed reduction of the energy scales of magnetic excitations
remained puzzling and could not be explained by theory so far.
\cite{Ole07}

In recent years the effect of doping in several cubic vanadium oxides
systems such as La$_{1-x}$Sr$_x$VO$_3$, Pr$_{1-x}$Ca$_x$VO$_3$,
Nd$_{1-x}$Sr$_x$VO$_3$, and Y$_{1-x}$Ca$_x$VO$_3$ has been
systematically explored by various experimental techniques.
\cite{Kas93,Pen99,Miy00,Fuj06,Fuj05,Fuj08,Sag08} Resistivity data,
specific heat and magnetic measurements have been used to set up
the phase diagram as function of doping.\cite{Fuj05} In contrast
to the high-$T_c$ cuprates, where the metal-insulator (MI)
transition is found at a few percent doping,\cite{Che91} in the
vanadates the MI transition occurs at much higher doping
concentrations:\cite{Dou75,Mot74,Kas93,Miy00} 18\% Sr in
La$_{1-x}$Sr$_x$VO$_3$ and even up to 50 \% Ca in
Y$_{1-x}$Ca$_x$VO$_3$. The evolution of optical spectra with
doping for these two systems shows that the defects lead to
impurity states which appear as absorption band deep inside the
Mott gap.\cite{Fuj08} This suggests that bound small
polarons are the cause of the MI transition at such high doping
concentrations.\cite{Fuj08} It is eventually the growth of the
mid-infrared absorption with increasing doping and the gradual
shift of this absorption toward zero energy which accompanies the
insulator-metal transition.

When taking all these experimental features into account
one arrives at a clear physical picture:\cite{Fuj08}
(i) most importantly, the trends of the optical conductivity show that
the edge of the Mott-Hubbard gap is essentially unaffected by doping,
and only fades away when the MI transition is approached;
(ii) defects play a central role, not just by introducing holes, but
as generators of deep impurity states which appear in the optical
conductivity as midgap absorption at low doping; and finally
(iii) the defects introduce two distinct energy scales.
On one hand one finds the defect
binding energy of about 1 eV, i.e., in the dilute doping regime,
and on the other hand there is clearly an activation energy of 
$\sim 0.1$ eV or less in transport experiments.\cite{Sag08,Dou75}
Our aim here is to show how generic Ca defects doped into the
Mott-insulator YVO$_3$ explain these phenomena in a natural way.

We begin our investigation with a discussion of the properties of
Ca defects inserted into an orbital degenerate Mott-Hubbard
insulator. We adopt a multiband Hubbard model description of the
$t_{2g}$ electrons,\cite{Ole05,Bou09} and we use the unrestricted 
Hartree-Fock (HF) method.\cite{Faz99,Wen10} This approach is chosen as
it allows us to introduce in a straightforward manner the
lower Hubbard band (LHB) of the $t_{2g}$ orbital states as well as the
upper Hubbard bands (UHBs) with the appropriate multiplet splitting.
The most pronounced effect of the defect is the Coulomb potential of 
the Ca-impurity which gives rise to an upward shift of the vanadium 
$t_{2g}$ states in the neighborhood of the defect. This leads to 
deep impurity states in the Mott-Hubbard gap. Interestingly, on one 
hand a Ca defect introduces one hole, but on the other hand it 
generates defect states on eight equivalent vanadium neighbors. 
\cite{notefn} Thus the topmost defect states that are split off
the LHB are partially filled and pin the chemical potential.

Transport inside the narrow defect band at weak doping will be affected
by the Coulomb disorder\cite{Che09} of the charged defects.
Consequently one expects Anderson localization as was actually already
conjectured by Mott\cite{Mot74} for the La$_{1-x}$Sr$_x$VO$_3$ system.
Subsequently we proceed to our central aim, namely the investigation of
the effect of defects on the spin-orbital dynamics in the dilute limit.
Hence important aspects concerning defects, namely:
(i) consequences of disorder,
(ii) the role played by the long-range Coulomb interaction,
and most importantly
(iii) the interaction effects at higher doping, will not be discussed
here, but will be addressed elsewhere.

One of the striking differences between the $G$-AF and $C$-AF
phase is the stability of the latter phase in doped
La$_{1-x}$Sr$_x$VO$_3$ and Y$_{1-x}$Ca$_x$VO$_3$ compounds. For
example, in  La$_{1-x}$Sr$_x$VO$_3$ the $C$-AF order survives even
beyond the insulator-to-metal transition at doping $x\simeq
0.18$,\cite{Fuj06} and disappears only at $x\simeq 0.26$.
\cite{Miy00,note26} The $G$-AF of YVO$_3$, however, is fragile and
is destabilized in the Y$_{1-x}$Ca$_x$VO$_3$ compounds already at
$x\simeq 0.02$, where the $C$-AF phase takes over.
\cite{Fuj05,Fuj08} We shall argue below that in order to
understand this behavior it is crucial to treat explicitly the
dynamics of orbital degrees of freedom. The cubic symmetry is
broken at V$^{3+}$ ions due to the orthorhombic lattice distortion
which occurs below the structural transition and favors
energetically the electron occupancy of $xy$
orbital.\cite{Bla01,Miy06} This symmetry breaking was also
confirmed by the electronic structure calculations performed for
LaVO$_3$ and YVO$_3$.\cite{Saw96,Sol06,And07} Due to Hund's
exchange both electrons in a $d^2$ configuration at V ion form a
high-spin $S=1$ state, so the second $t_{2g}$ electron occupies
either $yz$ or $xz$ orbital, resulting in a $xy^1(yz/zx)^1$ local
configuration at each V$^{3+}$ site.

The orbital state is quite different in both magnetic phases of 
YVO$_3$, below and above $T_{\rm N2}$. Lattice distortions are large 
in the low-temperature $G$-AF phase and suggest $C$-AO order.
This orbital order is further stabilized by increasing
pressure.\cite{Biz08} Above $T_{\rm N2}$ the distortions decrease
and are compatible with a weak $G$-type AO ($G$-AO)
order.\cite{Bla01} It was suggested by Ishihara \cite{Ish05} that
the phase transition at $T_{\rm N2}$ could originate from the
orbital physics and would be triggered by orbiton softening
induced by the reduction of the spin order parameter. While the
orbital degrees of freedom certainly play a role as the orbital
order indeed changes at $T_{N2}$, there is no evidence of orbiton
softening so far. It could be expected that the observed
transition is caused instead by local phenomena close to Ca
impurities in Y$_{1-x}$Ca$_x$VO$_3$ rather than by the global
change of orbital excitation scale. We suggest that it is
plausible that impurities could locally destabilize the $C$-AO
order, and introduce a microscopic model to treat this effect
below. An earlier theoretical analysis within the charge-transfer
model has shown that both phases are indeed energetically
close,\cite{Miz99} and one may thus expect that small changes of
the thermodynamic potential around $T_{N2}$ could trigger a first
order magnetic phase transition.

\begin{figure}[t!]
\includegraphics[width=7.2cm]{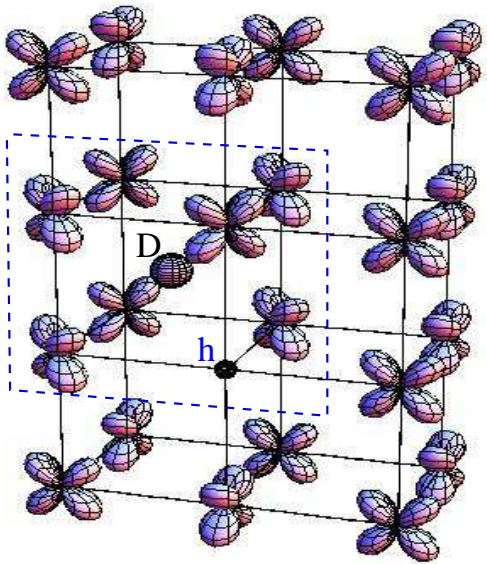}
\caption{(Color online) Schematic view of the lattice of vanadium
sites with occupied $t_{2g}$ orbitals and a single hole $h$
introduced by doping a Ca defect $D$ (sphere in the center) in the
$C$-type orbital structure of the low-temperature phase in weakly
doped Y$_{1-x}$Ca$_x$VO$_3$. For clarity the Y and O ions and the
orbital phases are not displayed. The hole occupies preferentially
one of the V sites that form a cube around the Ca impurity due to
the attractive Coulomb potential of the impurity. At the undoped V
sites only the topmost occupied $t_{2g}$ orbitals are shown. The
$|c\rangle\equiv|xy\rangle$ orbitals occupied at each V ion and
residing at lower energy are not shown. Orbital polarization, Eq.
(\ref{HD}), distorts the $|a\rangle\equiv|yz\rangle$
($|b\rangle\equiv|xz\rangle$) C-type alternating orbital order in
the neighborhood of the defect (dashed box) and favors occupation
of one of the two $\{|+\rangle_i,|-\rangle_i\}$ rotated orbital
states that minimize the orbital-defect interaction (see
Sec. \ref{sec:opa}). }
\label{fig:cao}
\end{figure}

The purpose of this paper is to investigate the local changes in the
electronic structure caused by Ca defects in the sublattice of Y ions,
and to derive the microscopic model leading to a scenario which 
could explain why the $G$-AF phase is so fragile under Ca doping in
Y$_{1-x}$Ca$_x$VO$_3$. Optical spectroscopy has shown that the
absorption in the Mott-Hubbard gap grows as a function of Ca doping in
Y$_{1-x}$Ca$_x$VO$_3$.\cite{Kas93,Fuj08} In the
low doping range $x\le 0.05$ it has a characteristic energy close to
1.2 eV,\cite{Fuj08} and  moves to lower energies at higher doping.
The intensity of the mid-infrared peak increases proportionally to
doping $x$. This new excitation energy which occurs in the doped system 
provides a measure of the binding energy of a doped hole which confines 
the hole to the V$_8$ cube in the immediate neighborhood of the charge 
defect, see Fig. \ref{fig:cao}. For the undoped bonds of this cube we 
invoke the mechanism of local polarization of orbitals near the 
Ca$^{2+}$ charge defects in the Y$^{3+}$ background. It will be shown 
that such defects, together with holes introduced into the $t_{2g}$
orbitals of vanadium ions, trigger the collapse of the $G$-AF order.

The major part of our study is based on an effective low-energy
Hamiltonian which can be denoted as a {\it spin-orbital\/} $t$-$J$ 
{\it model\/}; the model provides a description of magnetism and hole 
motion in the lower Hubbard band of a doped and orbital degenerate
Mott-Hubbard insulator. In the undoped case the model coincides
with the spin-orbital superexchange model
\cite{Kha01,Hor03,Kha04,Ole07,Hor08} which was shown earlier to
lead to a satisfactory description of spin-orbital physics, e.g.,
in YVO$_3$. The  spin-orbital $t$-$J$ model provides moreover the
description of the kinetic energy of doped holes and describes the
effect of charge impurities on the holes and the spin-orbital
degrees of freedom. In the concrete case of Y$_{1-x}$Ca$_x$VO$_3$
we show that the changes introduced by the defects have severe
effects on the orbital dynamics, namely: (i) are responsible for
the destabilization of the coexisting $G$-AF and $C$-AO order with
increasing doping $x$, and (ii) lead to an enhanced tendency
towards dimerization in the weakly doped $C$-AF phase.

The paper is organized as follows. First, in Sec. \ref{sec:Def},
we describe Ca$^{2+}$ charge defects in Y$_{1-x}$Ca$_x$VO$_3$
within the frame of a degenerate Hubbard model for $t_{2g}$
orbitals and analyze the effect of defects on the densities of
states derived within the HF approximation. In the
following Sec. \ref{sec:haf} the spin-orbital $t$-$J$ model is
outlined. The model Hamiltonian contains apart from spin-orbital
superexchange interactions also pure orbital interactions induced
by the lattice, see Sec. \ref{sec:sex}.
The effective double-exchange mechanism is introduced in Sec.
\ref{sec:hop}.
Next we derive the modified superexchange in the vicinity of the
hole in Sec. \ref{sec:sho} and develop the idea of orbital
polarization around charge defects in Y$_{1-x}$Ca$_x$VO$_3$ in
Sec. \ref{sec:opa}. With these terms, which supplement the
spin-orbital model for undoped cubic vanadates, the complete
$t$-$J$ model poses a rather complex many-body problem, and we
derive from it two effective one-dimensional (1D) embedded orbital 
models for the $G$-AF and $C$-AF phase separately, see Secs. 
\ref{sec:gaf} and \ref{sec:caf}. Thereby we treat spin correlations of 
$S=1$ spins in a classical approximation and focus on the orbital 
dynamics that plays a decisive role for the observed phase transition 
in Y$_{1-x}$Ca$_x$VO$_3$. Here we show how the orbital polarization
(see also Appendix A) develops around charge defects in the orbital 
chains for both magnetic phases. Next we consider a hole in both 
magnetic phases (Sec. \ref{sec:hole}), and derive the 1D orbital chain 
models appropriate for the $G$-AF phase and $C$-AF phase in Secs.
\ref{sec:de} and \ref{sec:holec}. Interactions around the hole are 
described in the $G$-AF phase by the $d^2-d^1$ superexchange 
(Appendix B) and the kinetic energy is controlled as in a typical 
double exchange mechanism by the spin orientations (discussed also in 
Appendix C). This implies, e.g., free hole hopping along the $c$ axis 
in the $C$-AF phase due to the FM alignment of spins. In Sec. 
\ref{sec:num} we introduce a statistical treatment of orbital 
correlations and energy contributions at finite doping (Sec. 
\ref{sec:sta}) and come to the conclusion that defects trigger 
dimerization of orbital correlations and of FM spin exchange constants
along the $c$ axis in the $C$-AF phase (Sec. \ref{sec:dim}). Finally, 
using the effective models derived for both magnetic phases we present 
a scenario which explains why the phase transition from the $G$-AF to
dimerized $C$-AF phase takes place already at small doping, see Sec. 
\ref{sec:pht}. A summary and outlook are given in Sec. \ref{sec:summa}.

\section{Defect states in $R$VO$_3$ perovskites}
\label{sec:Def}

\subsection{Degenerate Hubbard model for $t_{2g}$ electrons}
\label{sec:3-band}

We begin with analyzing the consequences of strong Coulomb
interactions in the framework of a multi-orbital Hubbard model
introduced here to describe the doped Y$_{1-x}$Ca$_x$VO$_3$ 
compounds. This model will later form the basis for the derivation 
of an effective superexchange and a related spin-orbital $t$-$J$
model for the orbital degenerate case, see Sec. \ref{sec:som}. In the 
ionic picture of the undoped YVO$_3$, V$^{3+}$ ions are in a $d^2$ 
electronic configuration with partly filled $t_{2g}$ orbitals. 
As we show below, the qualitative features of the optical spectra, 
i.e., the multiplet splitting of the Hubbard bands and the position of 
defect states, may be reproduced by considering a multiband Hubbard 
model for $t_{2g}$ electrons,
\begin{equation}
\label{hub}
H=H_{t}+H_{\rm CF}+H_{\rm int}+H_{\rm imp}\,,
\end{equation}
where the four terms stand for the kinetic energy ($H_{t}$),
crystal-field (CF) splitting ($H_{\rm CF}$), local Coulomb interactions 
($H_{\rm int}$), and Coulomb potential generated by a Ca
impurity ($H_{\rm imp}$). On one hand this model can also serve as
a basis for the qualitative discussion of the photoemission (PES)
and inverse PES in the vicinity of the Mott-Hubbard gap. On the
other hand, it provides the basis for the derivation of the
spin-orbital Hamiltonian which serves for a transparent
description of the magnetic and orbital structure, as well as of
spin and orbital excitations. Our aim here is to explore further
the changes of the excitation spectra resulting from the presence
of Ca$^{2+}$ defects in the lattice of Y$^{3+}$ ions and their
impact on the vanadium $t_{2g}$ electrons, with help of this
simplified Hamiltonian.

The kinetic energy is given by:
\begin{equation}
\label{ht} H_{t}=- \sum_{\gamma} \sum_{\langle
ij\rangle{\parallel}\gamma,\alpha(\gamma),\sigma}t_{\alpha}
\left(d^{\dagger}_{i\alpha\sigma}d^{}_{j\alpha\sigma}+
      d^{\dagger}_{j\alpha\sigma}d^{}_{i\alpha\sigma}\right),
\end{equation}
where $d^{\dagger}_{i\alpha\sigma}$ is electron creation operator
for an electron at site $i$ in orbital state $\alpha$ with spin
$\sigma=\uparrow,\downarrow$. The summation runs over three cubic
axes, $\gamma=a,b,c$, the bonds $\langle
ij\rangle{\parallel}\gamma$, and the hopping $t_{\alpha}$
conserves the $t_{2g}$ orbital flavor. The effective hopping
$t_{\alpha}$ originates from two subsequent $d-p$ hopping
processes via the intermediate $2p_{\pi}$ oxygen orbital along
each V--O--V bond. In principle it can be derived from the
charge-transfer model with $p-d$ hybridization $t_{pd}$ and
charge-transfer energy $\Delta$,\cite{Zaa93} and one expects in
the present case $t=t_{pd}^2/\Delta\sim 0.2$ eV.\cite{Kha01} Only
two $t_{2g}$ orbitals, labelled by $\alpha(\gamma)$, are active
along each bond $\langle ij\rangle{\parallel}\gamma$ and
contribute to the kinetic energy Eq. (\ref{ht}), while the third
one lies in the plane perpendicular to the $\gamma$ axis and the
hopping via the intermediate oxygen $2p_{\pi}$ oxygen is forbidden
by symmetry. This motivates a convenient notation used below,
\begin{equation}
\label{abc}
|a\rangle\equiv |yz\rangle, \hskip .7cm
|b\rangle\equiv |xz\rangle, \hskip .7cm
|c\rangle\equiv |xy\rangle,
\end{equation}
where the orbital inactive along a cubic direction $\gamma$ is
labelled by its index as $|\gamma\rangle$.

In agreement with a commonly accepted picture,\cite{Ren00} information
obtained from the electronic structure calculations,
\cite{Saw96,Sol06,And07} and with the results obtained using the point 
charge model,\cite{Hor08} we assume that the $xy$ ($c$) orbitals are 
energetically favored and thus occupied and inactive at low temperature,
while the remaining $yz$ and $xz$ orbitals are nearly degenerate and 
represent the $t_{2g}$ orbital doublet, with both orbitals active for 
the hopping (and the superexchange) along the $c$ cubic axis.

The nonequivalence of the $t_{2g}$ orbital states is described by
a CF splitting term which favors the $c$ orbitals,
\begin{equation}
\label{hcf}
H_{\rm CF}\equiv -\sum_{i\alpha\sigma}\Delta_{\alpha} n_{i\alpha\sigma},
\end{equation}
where $n_{i\alpha\sigma}\equiv
d^{\dagger}_{i\alpha\sigma}d^{}_{i\alpha\sigma}$ is an electron
density operator, and $\Delta_{\alpha}=\delta_{\alpha,c}\Delta_c$
with $\Delta_c>0$ . As a result, when $\Delta_c\geq t$ as we
estimated,\cite{Hor08} $c$ orbitals are filled by one electron at
each site in a strongly correlated system, and the second electron
occupies one of the orbitals in the $\{a,b\}$ doublet, leading to
the $c^1_i(a,b)^1_i$ configuration at each site $i$. This broken
symmetry situation corresponds to electron densities
\begin{equation}
\label{frozen}
n_{ic}\simeq 1, \hskip .7cm  n_{ia}+n_{ib}\simeq 1,
\end{equation}
within $t_{2g}$ orbitals
at undoped V$^{3+}$ ions. Note that one expects that the cubic symmetry
with $n_{i\gamma}=2/3$ is restored at high temperature, but this
situation will not be analyzed here as it has no influence on the
mechanism of the phase transition from the $G$-AF to $C$-AF phase which
occurs in Y$_{1-x}$Ca$_x$VO$_3$ at low temperature under increasing
doping.

The local Coulomb interactions between $t_{2g}$ electrons at V$^{3+}$
ions are described by the degenerate Hubbard Hamiltonian,\cite{Ole83}
with the interacting part:
\begin{eqnarray}
\label{Hee}
H_{\rm int}&=&
   U\sum_{i\alpha}n_{i\alpha  \uparrow}n_{i\alpha\downarrow}
 +\left(U-\frac{5}{2}J_H\right)\sum_{i,\alpha<\beta}n_{i\alpha}n_{i\beta}
\nonumber \\
&+& J_H\sum_{i,\alpha<\beta}
\left( d^{\dagger}_{i\alpha\uparrow}d^{\dagger}_{i\alpha\downarrow}
      d^{       }_{i\beta\downarrow}d^{       }_{i\beta\uparrow}
     +d^{\dagger}_{i\beta\uparrow}d^{\dagger}_{i\beta\downarrow}
      d^{    }_{i\alpha\downarrow}d^{       }_{i\alpha\uparrow}\right)
\nonumber \\
&-&2J_H\sum_{i,\alpha<\beta}\vec{S}_{i\alpha}\cdot\vec{S}_{i\beta}\,.
\end{eqnarray}
Here $n_{i\alpha}=\sum_{\sigma}n_{i\alpha\sigma}$ is the
corresponding electron density operator in orbital $\alpha$ at
site $i$, and spin operators ${\vec
S}_{i\alpha}=\{S_{i\alpha}^x,S_{i\alpha}^y,S_{i\alpha}^z\}$ are
related to fermion operators in the standard way, i.e.,
\begin{equation}
\label{S+z}
S_{i\alpha}^+\equiv d^{\dagger}_{i\alpha\uparrow}d^{}_{i\alpha\downarrow}\,,
\hskip .7cm
S_{i\alpha}^z\equiv\frac12(n_{i\alpha\uparrow}-n_{i\alpha\downarrow})\,.
\end{equation}
The first term in Eq.~(\ref{Hee}) describes the intraorbital
Coulomb interaction $U$ between electrons with antiparallel spins.
The second term stands for the interorbital Coulomb (density)
interaction, the third one is called frequently the "pair-hopping"
term, and the last one is Hund's exchange $J_H$. The choice of
coefficients in Eq. (\ref{Hee}) guarantees that the interactions
satisfy the rotational invariance in the orbital
space.\cite{Ole83} This Hamiltonian is exact when it describes
only one type of $3d$ orbitals which are partly occupied, as
$t_{2g}$ orbitals in the present case of the $R$VO$_3$ perovskites, 
and the interactions are then given by two parameters:
($i$) the intraorbital Coulomb element $U$ and ($ii$) the
interorbital (Hund's) exchange element $J_H$. These elements may
be expressed by the Racah parameters $\{A,B,C\}$. For $t_{2g}$
electrons one finds:\cite{Ole05,Gri71}
\begin{eqnarray}
\label{u}
U&=&A+4B+3C\,, \\
\label{jh}
J_H&=&3B+C\,.
\end{eqnarray}

Finally we introduce the Coulomb interaction
between the $t_{2g}$ electrons at a
V-site $\vec{r}_i$ and the effectively negative charged defects\cite{note1}
with charge $Q_D=e$ at site $\vec{R}_n$,
\begin{equation}
\label{Himp}
H_{\rm imp} = \sum_{i\in{\cal C}_n} W(|\vec{r}_i-\vec{R}_n|) n_i\,,
\end{equation}
where $n_i=\sum_{\alpha}n_{i\alpha}$ is the total $t_{2g}$
electron density. Here $i\in{\cal C}_1$ ($i\in{\cal C}_{\infty}$)
denotes a calculation where in the sum only nearest (all) V
neighbors of a defect are considered. The Coulomb potential itself
is long-ranged,
\begin{equation}
\label{Wimp}
W(r)=\frac{e Q_D}{\epsilon_c r},
\end{equation}
and screened by the
dielectric function $\epsilon_c$ of core electrons. We
identify $\epsilon_c$ with the high frequency dielectric constant
which, e.g., for YVO$_3$ lies in the interval $\epsilon(\omega)=5.0\pm 0.3$
in the whole frequency range $0.8<\omega<3.5$ eV.\cite{Tsv04}
The most pronounced effect of this potential term is an upward shift of
the vanadium states in the neighborhood of the defects.
This gives rise to bound states in the Mott-Hubbard gap. It
is important to realize that on one hand a Ca-defect $D$ introduces one
hole, but on the other hand it generates defect states on eight equivalent
vanadium neighbors. Thus the topmost defect states that are
split from the LHB are only partially filled and pin the chemical
potential $\mu$.

In general the coordinates of the defects {$\vec{R_n}$} will be
statistically distributed. In fact, defects will also feel some
repulsion and avoid clustering. We will not explore these aspects here,
as we are concerned mainly with the dilute doping regime. Nevertheless
it is clear that transport is strongly affected by disorder and results
from the motion of holes in the defect band.

Furthermore, in the immediate neighborhood of a Ca defect the
strength of Coulomb interaction influences the orientation of $t_{2g}$
orbitals filled by electrons. This effect modifies the orbital state
and will be described below via an extra crystal field term which acts 
on the orbitals of the V ions in the vicinity of a defect (see Sec. 
\ref{sec:opa}). Such more subtle effects concern the low-energy 
behavior and are analyzed in the framework of the effective spin-orbital 
$t$-$J$ model in Sec. \ref{sec:haf}. Therefore, we do not analyze it 
here, similar as the intersite orbital interactions which originate 
from the distortions of VO$_6$ octahedra.

\subsection{ Unrestricted Hartree-Fock analysis
           of the degenerate Hubbard model }
\label{sec:uhf}

Qualitative insight into the correlated electronic structure of partly 
filled $t_{2g}$ states can be obtained from the HF approximation. When
the HF approximation is used, the "pair-hopping" term does not 
contribute and the local spin exchange interactions 
${\vec S}_{i\alpha}\cdot{\vec S}_{i\beta}$ contribute only with the 
Ising term $\propto S_i^zS_j^z$,\cite{Fle97} i.e., 
one may use an approximate expression,
\begin{equation}
\label{Hund} -2{\vec S}_{i\alpha}\cdot{\vec
S}_{i\beta}\simeq\frac{1}{2} \sum_{\sigma}\left(
n_{i\alpha\sigma}n_{i\beta{\bar{\sigma}}} -
n_{i\alpha\sigma}n_{i\beta\sigma}\right)\,.
\end{equation}
As a result, local electron interactions Eq. (\ref{Hee}) are given
approximately by electron density operators
$\{n_{i\alpha\sigma}\}$:
\begin{eqnarray}
\label{Happ}
H_{\rm int}&\simeq&
   U\sum_{i\alpha}n_{i\alpha  \uparrow}n_{i\alpha\downarrow}
+(U-2J_H)\sum_{i,\alpha<\beta,\sigma} n_{i\alpha\sigma}n_{i\beta{\bar{\sigma}}}
\nonumber\\
&+&(U-3J_H)\sum_{i,\alpha<\beta,\sigma} n_{i\alpha\sigma}n_{i\beta\sigma}\,.
\end{eqnarray}
The form given in Eq. (\ref{Happ}) violates the rotational invariance 
of local Coulomb interactions,\cite{Ole83} but is sufficient for our 
purpose since all the terms which appear in the HF approximation arise 
from it.\cite{Fle97} One finds the following effective
one-particle problem in an effective field,
\begin{eqnarray}
\label{HHF}
H_{\rm int}^{\rm HF}\!&=&\!
U\sum_{i\alpha\sigma}n_{i\alpha\sigma}\langle n_{i\alpha{\bar{\sigma}}}\rangle
\nonumber\\
\!&+&\!(U-2J_H)\sum_{i,\alpha<\beta,\sigma}\left(
 n_{i\alpha\sigma}\langle n_{i\beta{\bar{\sigma}}}\rangle
+\langle n_{i\alpha\sigma}\rangle n_{i\beta{\bar{\sigma}}}\right)
\nonumber\\
\!&+&\!(U-3J_H)\sum_{i,\alpha<\beta,\sigma}\left(
n_{i\alpha\sigma}\langle n_{i\beta\sigma}\rangle
+\langle n_{i\alpha\sigma}\rangle n_{i\beta\sigma}\right)\,.
\nonumber\\
\end{eqnarray}
The HF potentials are determined by the average
densities $\{\langle n_{i\alpha\sigma}\rangle\}$ when $H_{\rm
int}$ is replaced by $H_{\rm int}^{\rm HF}$ in Eq. (\ref{Happ}).

We emphasize that although the quantum effects such as spin
fluctuations and "pair hopping" of double occupancies are
neglected, the essential features of the Coulomb interaction are
reproduced. This can be seen by considering charge excitations
$d^n_id^n_j\rightarrow d^{(n+1)}_id^{(n-1)}_j$ along a given bond
$\langle ij\rangle$ in the lattice of transition metal ions with
$d^n$ electronic configuration. To analyze excited states we
assume that the electron number $n$ is lower than the half-filled
shell, i.e., $n<5$ for the $3d$ shell (below we focus on $n=2$ for
the present problem of V$^{3+}$ ions in YVO$_3$, where the
$t_{2g}$ shell is half-filled at $n=3$). There are three types of
$d^{(n+1)}$ excited states: (i) high-spin (HS) states with all
electrons in the same spin state (realized for a FM $\langle
ij\rangle$ bond); (ii) low-spin (LS) states with all orbitals
being singly occupied, and (iii) LS states with one doubly
occupied orbital. The energies of these excitations are:
\begin{eqnarray}
\label{ehs}
E_{\rm HS}&=&U-3J_H\,,
\\
\label{els1}
E_{\rm LS}^{(1)}&=&U-2J_H+(n-1)J_H\,,
\\
\label{els2}
E_{\rm LS}^{(2)}&=&U    +(n-1)J_H\,.,
\end{eqnarray}
The HS excitation energy obtained in the HF approximation reproduces 
the exact value obtained by the diagonalization of the atomic ion
Hamiltonian Eq. (\ref{Hee}). The remaining energies are
systematically lower by $J_H$ than the exact values:\cite{notex}
\begin{eqnarray}
\label{els1ex}
E_{\rm LS,ex}^{(1)}&=&U-J_H+(n-1)J_H\,,
\\
\label{els2ex}
E_{\rm LS,ex}^{(2)}&=&U+J_H+(n-1)J_H\,,
\end{eqnarray}
as the quantum fluctuation effects (spin-flips and "pair-hoping")
were neglected.

\begin{figure}[t!]
\includegraphics[width=8.2cm]{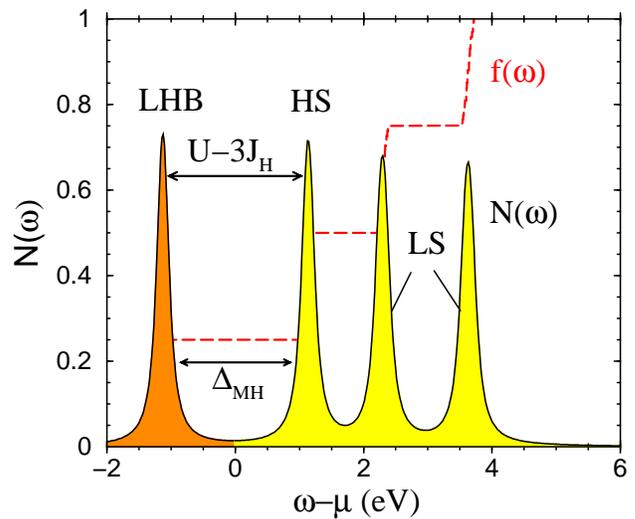}
\caption{(Color online) Partial density of states $N(\omega)$ (in
eV$^{-1}$ units) for $t_{2g}$ states with $\{yz,zx\}$ symmetry of
the Mott insulator YVO$_3$ obtained by the unrestricted
HF calculation. The multiplets represent the occupied
lower Hubbard band (LHB) while the unoccupied upper Hubbard band
consists of high-spin (HS) and low-spin (LS) multiplet states,
respectively. The Mott-Hubbard gap $\Delta_{\rm MH}$ separates the
LHB and the HS band. The dashed step-like curve indicates the
integrated number of states $f(\omega)$ Eq. (\ref{f}) normalized
to one. Parameters: $U=4$, $J_H=0.6$, $t_c=t_a=t_b=0.2$ (all in
eV). }
\label{fig:N_w_8_0}
\end{figure}

In the relevant regime of parameters for the $R$VO$_3$
perovskites, the electrons in the undoped YVO$_3$ are localized in
a Mott insulator. We consider here representative parameters,
\cite{Ole07} with $U=4$ eV and $J_H=0.6$ eV. In this parameter range 
one may simplify the problem of calculating the electronic structure 
in the HF approximation as the $c$ orbitals are occupied
at each site by one $t_{2g}$ electron, and the magnetic state is
determined by their magnetic moments. This follows from large
Hund's exchange which dictates that the spin direction of both
$t_{2g}$ electrons ($c$ electron and $(a,b)$ one) agree at each
site. As a result, one obtains the high-spin $S=1$ state at each
V$^{3+}$ ion.

We discuss first the numerical calculations of the density of
states $N(\omega)$ for the undoped system. The results were obtained 
by solving the equations derived using the HF approximation, see Eq.
(\ref{HHF}), on a cluster with $N_0=L_x\times L_y\times L_z$ sites
(here we use $L_x=L_y=L_z=10$). We considered the $c$ orbitals
singly occupied and ordered as in the $C$-AF phase of the undoped
YVO$_3$. The second electron at each site occupies one of the
remaining $t_{2g}$ orbitals $\{a,b\}$. The partial density of
states for the $\{a,b\}$ orbital doublet with eigenenergies
$\varepsilon_m$ in the cluster,
\begin{equation}
\label{Nw} N(\omega)=\sum_m\delta(\omega-\varepsilon_m) ,
\end{equation}
is shown in Fig. \ref{fig:N_w_8_0}. The chemical potential $\mu$
in the undoped case is naturally chosen in the middle of the gap:
$\mu=(\epsilon_{N_0}+\epsilon_{N_0+1})/2$, where $\epsilon_{N_0}$
is the highest occupied energy for $N_0$ electrons in $\{a,b\}$
orbitals in the system.

\begin{table}[t!]
\caption{ Excitation energies of YVO$_3$ inferred from optical
data\cite{Fuj08,Kue09} and calculated from the present theory.
Here $\Delta_{\rm MH}$ is the Mott-Hubbard gap, and $E_{\rm HS}$,
$E_{\rm LS}^{(1)}$ and $E_{\rm LS}^{(2)}$ are the energies
(all in eV) of the high-spin and the two low-spin transitions
measured from the center of the lower Hubbard band $E_{\rm LHB}$.
The last column gives the defect energy $E_D$ relative to $E_{\rm
LHB}$ at $x=0.02$ Ca-doping. } \vskip .2cm
\begin{ruledtabular}
\begin{tabular}{cccccccc}
& energy & $\Delta_{\rm MH}$ & $E_{\rm HS}$ & $E_{\rm LS}^{(1)}$
                             & $E_{\rm LS}^{(2)}$& $E_D$ & \cr
\colrule
& Ref. \onlinecite{Kue09} &  1.7  &  $2.2$  &  3.0--3.7  & 4.0--4.5 &  ---  &  \cr
& Ref. \onlinecite{Fuj08} &  1.7  &  $2.2$  &  ---  &  ---  &  1.2  &  \cr
\colrule
& theory                  &  1.6  &   2.2   &  3.4  &  4.6  &  1.0  &  \cr
\end{tabular}
\end{ruledtabular}
\label{tab:uj}
\end{table}

The spectra are characterized by four subbands: (i) the LHB
centered at $E_{\rm LHB}$ and (ii) the UHB which itself is split
into three subbands (multiplet structure) corresponding to the HS
excitations at $E_{\rm HS}=U-3J_H$ Eq. (\ref{ehs}), and two LS
transitions centered at $E_{\rm LS}^{(1)}=U-J_H$ Eq. (\ref{els1})
and $E_{\rm LS}^{(2)}=U+J_H$ Eq. (\ref{els1}), respectively. These
energies are relative to $E_{\rm LHB}$. HF results for
these excitations calculated for $U=4$, $J_H=0.6$ and $t_0=0.2$ eV
($t_a=t_b=t_c=t_0$) are listed in Table I and compared with
experimental values for YVO$_3$ obtained by Fujioka {\it et al.\/}
\cite{Fuj08} and by K\"upersbusch\cite{Kue09} deduced from optical
spectroscopy and ellipsometry, respectively. Interestingly the
LS-transitions found in the ellipsometry study find a reasonable
correspondence within the HF calculation. These values are also
consistent with the underlying multiplet splitting of  YVO$_3$ in
the optical spectral weight study of Ref. \onlinecite{Kha04}.

Next we turn to the Mott-Hubbard gap (see Fig. \ref{fig:N_w_8_0})
which may be expressed as:
\begin{equation}
\label{Mottgap} \Delta_{\rm MH}\simeq U-3 J_H-W_{\rm eff},
\end{equation}
where $W_{\rm eff}$ is the effective bandwidth of a Hubbard
subband. A significant reduction of $W_{\rm eff}$ compared to the
free bandwidth $W$ is well known from the single band Hubbard
model.\cite{Ste90} In a recent study a LHB width  $W_{\rm
eff}\approx 3 t_0$ was estimated for the incoherent hole motion of
$t_{2g}$ electrons.\cite{Woh09} Thus with the choice $t_0\simeq
0.2$ eV we obtain for the Mott-Hubbard gap $\Delta_{\rm MH}\simeq
1.6$. We also note, that our estimate of the bandwidth of the LHB
in the HF calculation is much smaller, namely $W_{\rm
eff}^{\rm HF}\approx t_0$. This is due to the neglect in the
HF approximation of processes describing the incoherent motion of
holes. A related interesting quantity that can be inferred from
optical spectroscopy is the width of the HS-transition,\cite{Fuj08} 
$\Gamma_{\rm HS}\simeq 1.3\pm 0.2$ eV in YVO$_3$. As the width of the 
optical transition should be determined by the convolution of the LHB 
and HS-band, one expects $\Gamma_{\rm HS}\simeq 2 W_{\rm eff}$. 
Thus $\Delta_{\rm MH}$ and the width $\Gamma_{\rm HS}$ find a natural 
explanation in terms of the multiplet splitting and the effective 
Hubbard bandwidth $W_{\rm eff}$.

\begin{figure}[t!]
\includegraphics[width=7.5cm]{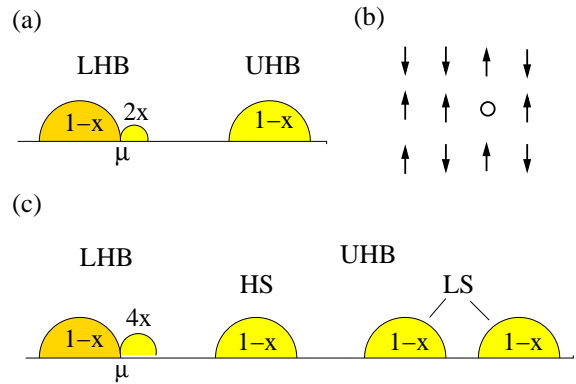}
\caption{(Color online) Spectral weight of Hubbard subbands per
site at hole doping $x$ for: (a) the single band Hubbard model and
(c) Hubbard bands of doubly degenerate model with two spin and two
orbital flavors. At doping concentration $x$ the number of states
$(1-x)$ in each subband of the UHB is determined by the number of
filled states in the LHB. Whereas in (a) a hole can be filled
either by an up- or a down-spin electron, as shown for a
representative electron configuration in (b), in the orbital
degenerate case (c) there are 4 choices to fill up the hole in the
atomic limit, i.e., $4x$ empty states in the LHB for a system with
$x$ doped holes. } \label{fig:weight}
\end{figure}

The counting of states that contribute to the partial density of states
$N(\omega)$ is straightforward; there is one electron per site
which can occupy either $a$ or $b$ orbital and there are
two spin flavors. Thus the filling is $f(\mu)=1/4$ as the total number
of states is $N_t=4 N_0$, where $N_0$ is the number of sites, and there
are $N_0$ electrons that occupy the LHB. All subbands of the UHB have
also the same weight of $N_0$ each.
This can be seen from the integrated and normalized density of states
\begin{equation}
\label{f} f(\omega)=\int\limits_{-\infty}^{+\omega}d{\omega}'N(\omega')
/\int\limits_{-\infty}^{+\infty}d{\omega}'N(\omega') ,
\end{equation}
which is displayed as the dashed curve in Fig. \ref{fig:N_w_8_0}.
While we use a broadening parameter $\gamma=0.1$ eV
to smoothen the density of states $N(\omega)$, no such broadening
is used in the calculation of $f(\omega)$.

It is well known that the number of states in the UHB depends on
the hole doping,\cite{Uch91,Ste90} i.e., varies with electron
filling in striking contrast to bands of ordinary semiconductors
or insulators. It has also been realized that by creating a hole
in a Mott insulator (with no orbital degeneracy) actually two
empty states are created in the LHB, see Fig. \ref{fig:weight}(a)
as there are two options to fill up the
hole.\cite{Mei93,Esk94,Phi10} That is, at a concentration of $x$
holes, there is a fraction of $2x$ unoccupied states in the LHB
above the chemical potential, and simultaneously the number of
states in the UHB is reduced by a factor $1-x$. Such a spectral
weight transfer upon electron doping has been recently observed in
TiOCl.\cite{Sin11}

In case of the present model with two orbital flavors $\{a,b\}$ in
addition to spin, there is one LHB and three UHBs, all of them
with the same weight one at $x=0$. At finite doping $x>0$ each
added hole moves an empty state above the Fermi energy in the LHB
and generates also three other unoccupied states in the LHB which
are taken from the subbands of the UHB, see
Fig.\ref{fig:weight}(c). This corresponds to four possibilities to
fill up a hole (with spin and orbital flavor) and thus $4x$ empty
states belong to the LHB at doping $x$. Yet it is also important
to emphasize here that this does not imply that there are really
$4x$ free states in the LHB that can all be simultaneously
occupied. Actually there is only space for $x$ electrons, as with
each electron added three states are shifted back to the UHBs.

\subsection{Hartree-Fock calculation for defect states}
\label{sec:rel}

\begin{figure}[t!]
\includegraphics[width=8.2cm]{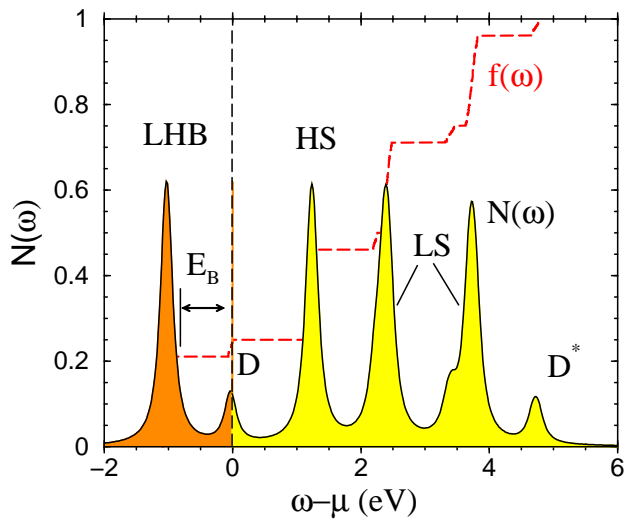}
\caption{(Color online) Partial density of states of a doped Mott
insulator with degenerate $t_{2g}$ orbitals $\{a,b\}$ at $x=0.02$
doping concentration. Partially filled defect states $D$ are split
off from the LHB by the defect potential $V_D=1$ eV. The vertical
dashed line indicates the filling of the system at finite doping;
$D^*$ marks defect states split from the highest LS excitation;
further notations and parameters as in Fig. 2.} \label{fig:Nw8}
\end{figure}

The major change in the spectra induced by doping is the occurrence of 
the defect states $D$ inside the Mott-Hubbard gap. Figure \ref{fig:Nw8} 
shows the density of states at 2\% Ca-doping. The spectra were obtained 
by a calculation using the HF approximation for well separated defects 
acting on $t_{2g}$ electrons with the defect potential, Eq. 
(\ref{Himp}), which includes the interaction with the eight nearest 
neighbor V ions at distance $d_1=\sqrt{3}\;d_{VO}$ 
($d_{\rm VO}\simeq 2.0$ \AA\; is the vanadium-oxygen distance). 
The estimate of the nearest neighbor
defect potential $V_D$ appropriate for YVO$_3$ is straightforward,
i.e., $W(d_1)=V_D\simeq 1.0$ eV, where we used the dielectric
constant of core electrons $\epsilon_c\simeq 5.0$.
The summation in Eq. (\ref{Himp}) was restricted to $i\in{\cal C}_1$,
thus only a short-range defect potential was included, and disorder
effects do not come into play. Thereby the interpretation of spectra
and energy scales
is simplified and more transparent for our purposes.

\begin{figure}[t!]
\includegraphics[width=8.2cm]{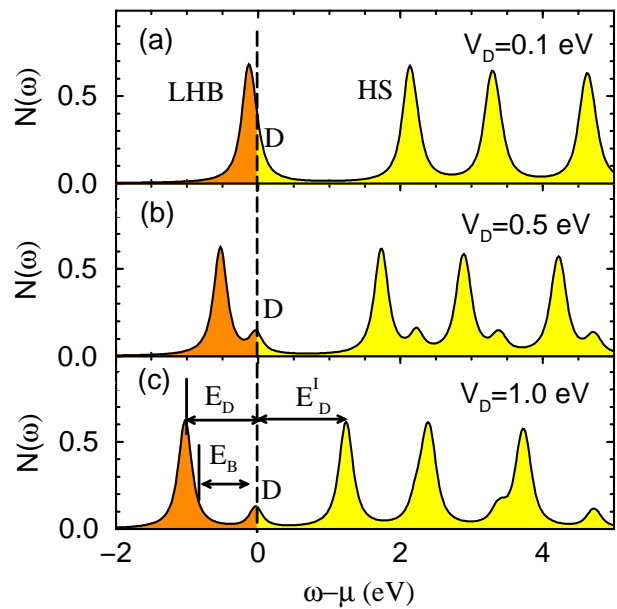}
\caption{(Color online) Partial density of states $N(\omega)$ of a
$t_{2g}$ Mott insulator at $x=0.02$ doping concentration for
different values for the defect potential: (a) $V_D=0.1$ eV, (b)
$V_D=0.5$ eV, and (c) $V_D=1.0$ eV. The vertical dashed line
indicates the chemical potential $\mu$ which is pinned by the
defect states $D$. The binding energy $E_B$ of the defect states
measured from the edge of the LHB grows with increasing $V_D$. An
alternative measure is the defect energy $E_D$ ($E'_D$) relative
to the center of the LHB (HS-band), respectively. Parameters:
$U=4$, $J_H=0.6$, $t_c=t_a=t_b=0.2$ (all in eV). }
\label{fig:Nw9}
\end{figure}

The calculations show that for large $V_D\simeq 1.0$ eV the energy
$E_D$ of the defect states $D$ relative to the center of the LHB
is approximately given by $E_D\approx V_D$ (see Table I). The
complementary splitting of the defect states from the center of
the HS-band is $E'_D\simeq 1.2$ eV, and appears also consistent
with optical spectroscopy data.\cite{Fuj08} Depending on the value
of the defect potential, the defect states $D$ appear either at
the upper edge of the LHB, or develop to a separated maximum
within the Mott-Hubbard gap, see Fig. \ref{fig:Nw9}. Figure
\ref{fig:Nw9}(b) nicely shows that each UHB (HS, LS$^{(1)}$ and
LS$^{(2)}$) has its own defect satellite. Such states can be
observed, however, only when a satellite of a given Hubbard
subband is well separated from the next subband.

We emphasize that in the model with two orbital flavors each
Ca$^{2+}$ defect introduces one hole into eight defect states that
are split off from the LHB. The filling fraction at doping
concentration $x=N_x/N_0$ is therefore,
\begin{equation}
\label{fd} f(\mu)=\frac{N_0-N_x}{4 N_0}=\frac{1}{4}\,(1-x)  ,
\end{equation}
which fixes the chemical potential $\mu$. The total number of occupied
electron states is now simply obtained by multiplication with $4 N_0$:
\begin{equation}
\label{nLHB} N^{\rm occ}=(1-x) N_0  ,
\end{equation}
and the number of holes is $N^{\rm h}=x N_0$. The number of states in 
the LHB, however, is reduced to $N_{\rm LHB}=(1-8x)N_0$ due to the 
appearance of $N_D=8x N_0$ defect states, of which 
$N^{\rm occ}_D=7x N_0$ are occupied.

Although there is a similarity to defect bands in doped
semiconductors, \cite{Mot74} a striking difference is that the
defect states here are completely derived from the LHB. Hence the
defect band in Fig. \ref{fig:Nw9}(c) would be fully occupied if no
holes were added. However each added defect generates not only the
defect states but also contributes a single hole. We suggest that
it is the motion of the holes in the defect band, i.e., via
hopping from one defect to the next one which occurs in presence
of Coulomb disorder that explains the small excitation energies
observed in transport experiments.\cite{Sag08,Dou75}

\section{Spin-orbital $t$-$J$ model }
\label{sec:haf}

\subsection{Model Hamiltonian for Y$_{1-x}$Ca$_x$VO$_3$ }
\label{sec:som}

We now turn to the derivation of an effective low-energy
Hamiltonian which describes the interactions of the spin and
orbital degrees of freedom as well as the motion of doped holes.
For the undoped case the effective microscopic spin-orbital model
that describes electrons in $t_{2g}$ orbitals of V ions has
already been worked out.\cite{Kha01,Hor03,Kha04,Ole07,Hor08} 
In the undoped compound one deals with the $t_{2g}^2$ electronic
high-spin ($S=1$) state at each V$^{3+}$ ion, and an $\{a,b\}$
orbital degree of freedom.

When a correlated insulator with active orbital degrees of freedom
is doped, rather complex processes occur when holes are doped. 
The motion of a hole may generate spin, orbital, or spin-orbital
excitations on its path.\cite{Zaa93,Zaa92,vdB00,Dag08,Woh09}
Therefore, designing a microscopic model which captures the
essential physical mechanisms in such a situation and is
mathematically tractable at the same time is a nontrivial and
demanding task. Below we introduce such a microscopic model to
describe the changes of magnetic and orbital order that occur by hole
doping in Y$_{1-x}$Ca$_x$VO$_3$. It stems from the spin-orbital model 
for the undoped $R$VO$_3$ perovskites and includes the superexchange 
and the orbital interactions induced by the lattice.\cite{Kha01,Ole07} 
As in Sec. II, the kinetic energy arises from the hopping $t$ between 
two $t_{2g}$ orbitals at neighboring V sites, and the orbital flavor
is conserved.\cite{Zaa93} Electron-electron interactions are
described by the degenerate Hubbard model\cite{Ole83} Eq.
(\ref{Hee}), written in this case for two $t_{2g}$ orbitals
$\{a,b\}$, with intraorbital Coulomb element $U$ and Hund's
exchange $J_H$, see Eqs. (\ref{u}) and (\ref{jh}).

As in other Mott (or charge-transfer) insulators characterized by
the realistic regime of parameters with $t\ll U$,\cite{Ole05} the
magnetic and orbital interactions of strongly correlated $t_{2g}$
electrons in the $R$VO$_3$ perovskites are then described within
the spin-orbital superexchange ${\cal H}_J$, with the
superexchange constant
\begin{equation}
\label{jex} J=\frac{4t^2}{U}\,.
\end{equation}
A realistic model for the undoped $R$VO$_3$ perovskites contains
also the orbital interactions ${\cal H}_{\rm orb}$ which follow
from the orbital-lattice coupling and are responsible both for the
energetic proximity of the $G$-AF and $C$-AF phases in YVO$_3$,
\cite{Kha01,Hor03,Ole07} and for the systematic trends observed for 
the orbital and magnetic phase transition of the series of $R$VO$_3$
perovskites.\cite{Hor08} Here we present an appropriate extension
of this model adequate for weakly doped Y$_{1-x}$Ca$_x$VO$_3$
compounds, which includes the charge-orbital interactions around
the Ca$^{2+}$ defects. It generates an attractive potential and
confines a hole to the immediate neighborhood of the charge
defect.

Strong electron correlations in the Y$_{1-x}$Ca$_x$VO$_3$ compounds
motivate the spin-orbital $t$-$J$ model,
\begin{equation}
\label{tJ} {\cal H}_{t_{2g}}={\cal P}\,\Big( {\cal H}_t+{\cal
H}_{\rm Hund}+{\cal H}_J+{\cal H}_{\rm orb}+{\cal H}_{\rm imp}
+{\cal H}_D+{\cal H}_I\Big)\,{\cal P}\,,
\end{equation}
where ${\cal P}$ are the projection operators which remove triply
(and higher) occupied V ions and guarantee that either the
$\{a,b\}$ orbital doublet is singly occupied (at V$^{3+}$ ion) and
spin is $S=1$, or the $\{a,b\}$ doublet is empty (at V$^{4+}$
ion), i.e., it contains a hole generated by doping. The
spin-orbital superexchange for $d^2-d^2$ pairs of
V$^{4+}$-V$^{3+}$ ions (${\cal H}_J$), and the orbital-lattice
interactions (${\cal H}_{\rm orb}$) stand for the effective strong
coupling model,\cite{Kha01} see Sec. \ref{sec:sex}, that was used 
before to explain the temperature variation of optical spectra,
\cite{Kha04} and the phase diagram\cite{Hor08} of the $R$VO$_3$ 
perovskites. When holes are doped, several other terms are needed: 
(i) the hopping of $\{a,b\}$ electrons in the restricted space 
(${\cal H}_t$) Eq. (\ref{ht}); (ii) Hund's exchange 
${\cal H}_{\rm Hund}$ between $s=1/2$ spins of an $\{a,b\}$ electron 
and a $c$ electron at site $i$; (iii) the impurity potential 
${\cal H}_{\rm imp}$ given by Eq. (\ref{Himp}); (iv) the polarization 
interaction at the V sites near the charge defects (${\cal H}_D$); 
(v) the superexchange for $d^2-d^1$ pairs of V$^{4+}$-V$^{3+}$ ions 
(${\cal H}_I$). These terms are introduced in Secs. \ref{sec:hop}, 
\ref{sec:opa} and \ref{sec:sho}, respectively.

\subsection{Superexchange in the undoped YVO$_3$}
\label{sec:sex}

The third term in Eq. (\ref{tJ}) is the spin-orbital superexchange
${\cal H}_J$. The anisotropic electron distribution between the
$ab$ planes and the $c$ axis, see Eqs. (\ref{frozen}), is
responsible for a particular form of the superexchange ${\cal
H}_J$, with broken cubic symmetry.\cite{Kha01,Hor03,Kha04} In
fact, this symmetry breaking is responsible for strong $\{a,b\}$
orbital fluctuations which stabilize the $C$-AF phase with FM
interactions along the $c$ axis. The superexchange interactions
between two V$^{3+}$ ions in the undoped YVO$_3$ with $S=1$ spins
at sites $i$ and $j$ arise from virtual excitations
$d^2_id^2_j\rightleftharpoons d^3_id^1_j$ along the concerned bond
$\langle ij\rangle$, promoted by the hopping $t$ which couples
pairs of identical {\it active $t_{2g}$ orbitals\/}. A single
hopping process generates a $d^3$ configuration at site $i$,
either with three different orbitals occupied by a single electron
each, or with a double occupancy in one of the two active
orbitals. Therefore, the $d^3_i$ excited state may be either a
high-spin $^4A_2$ state with energy $(U-3J_H)$, see Eq.
(\ref{ehs}), or one of three low-spin states: $^2E$, $^2T_1$ or
$^2T_2$ with energies:\cite{notex} $U$ and $(U+2J_H)$ --- they are
shown in Fig. 1 of Ref. \onlinecite{Ole05}, see also Eqs.
(\ref{els1ex}) and (\ref{els2ex}).

This perturbative consideration leads to the spin-orbital
superexchange model for $S=1$ spins in cubic vanadates,\cite{Kha01}
\begin{eqnarray}
\label{HJ} {\cal H}_{J}\!\!&=&\!\!\frac{1}{12}\,J\sum_{\langle
ij\rangle\parallel c} n_in_j\left\{ 4r_1 (2+\vec S_i\!\cdot\!\vec
S_j)
\left(\vec\tau_i\cdot\vec\tau_j-\frac{1}{4}\right)\right. \nonumber \\
&+&\!\!\left. (\vec S_i\!\cdot\!\vec S_j-1)
\left(\frac{7}{4}-\tau_i^z\tau_j^z-\tau_i^x\tau_j^x
+5\tau_i^y \tau_j^y\right)\right.                        \nonumber  \\
&+&\!\!\left. 3r_3 (\vec S_i\!\cdot\!\vec S_j-1)
\left(\frac{1}{4}+\tau_i^z\tau_j^z+\tau_i^x\tau_j^x
-\tau_i^y \tau_j^y\right)\right\}                        \nonumber  \\
&+&\!\!\frac{1}{24}\,J\sum_{\langle ij\rangle\parallel
ab}n_in_j\left\{ 4r_1\left(\vec S_i\!\cdot\!\vec S_j+2\right)
\left(\tau_i^z\tau_j^z-\frac{1}{4}\right)\right.         \nonumber  \\
&+&\!\!\left. 3(1-\vec S_i\!\cdot\!\vec S_j)
\left(\frac{19}{12}\mp \frac{1}{2}\tau_i^z
\mp \frac{1}{2}\tau_j^z-\frac{1}{3}\tau_i^z\tau_j^z\right)\right. \nonumber  \\
&+&\!\!\left. 3r_3(1-\vec S_i\!\cdot\!\vec S_j)
\left(\frac{5}{4}\mp \frac{1}{2}\tau_i^z \mp
\frac{1}{2}\tau_j^z+\tau_i^z\tau_j^z\right)\right\},
\end{eqnarray}
where $\langle ij\rangle$ is a nearest neighbor bond and $J$ is
the superexchange constant given in Eq. (\ref{jex}). The
superexchange Eq. (\ref{HJ}) follows from the degenerate Hubbard
model Eq. (\ref{hub}) as described in Ref. \onlinecite{Ole07} for
the case when $c$ orbitals are occupied, see Eq.
(\ref{frozen}), and a second electron occupies the $\{a,b\}$
doublet at each site, i.e., $n_i=1$ with
\begin{equation}
\label{ni} n_i\equiv n_{ia}+n_{ib}.
\end{equation}
It depends on Hund's exchange,
\begin{equation}
\label{eta} \eta=\frac{J_H}{U}\,,
\end{equation}
due to the charge excitations to the states of the UHBs described
above and shown in Fig. \ref{fig:N_w_8_0}, via the coefficients
(the coefficient $r_2=1$ arises for the intermediate energy LS
excitations at $V^{+2}$ ions),
\begin{equation}
\label{rr} r_1=\frac{1}{1-3\eta}\,, \hskip 1cm
r_3=\frac{1}{1+2\eta}\,.
\end{equation}
The operators $\vec\tau_i\equiv\{\tau_i^x,\tau_i^y,\tau_i^z\}$
describe orbital $\tau=1/2$ pseudospins defined here (for each
direction $\gamma=a,b,c$) by the doublet of active
$\{a,b\}\equiv\{yz,xz\}$ orbitals, and are given by the Pauli
matrices, i.e.,
\begin{equation}
\label{Pauli} \tau_i^x\equiv\frac12\,\sigma_i^x\,, \hskip .7cm
\tau_i^y\equiv\frac12\,\sigma_i^y\,, \hskip .7cm
\tau_i^z\equiv\frac12\,\sigma_i^z\,.
\end{equation}
As both orbitals are active along the $c$ axis, the orbital part
is then given by a scalar product $\vec\tau_i\cdot\vec\tau_j$.

The orbital-orbital interactions which follow from lattice
distortions of both the Jahn-Teller type and GdFeO$_3$-type are of
the form: \cite{Kha01,Hor03,Kha04,Ole07}
\begin{equation}
\label{HJT} {\cal H}_{\rm orb} =
 V_{ab}\sum_{\langle ij\rangle\parallel ab}\tau^z_i\tau^z_j
-V_c\sum_{\langle ij\rangle\parallel c}\tau^z_i\tau^z_j\,.
\end{equation}
The interactions in the ab planes $V_{ab}>0$ follow from the
Jahn-Teller distortions, while the ones along the $c$ axis $V_c>0$
favor the $C$-AO phase and thus help to stabilize the $G$-AF order
in the undoped YVO$_3$. These interactions increase along the
$R$VO$_3$ perovskites towards the compounds with small ionic
radii, and it happens to be just for YVO$_3$ that they tip the
balance between the two types of magnetic order in favor of the
$G$-AF phase which is more stable at low temperature.\cite{Miy06}

\subsection{Effective double exchange model}
\label{sec:hop}

The first term $\tilde{\cal H}_t\equiv {\cal P}{\cal H}_t{\cal P}$
in Eq. (\ref{tJ}) is the kinetic energy which after projection
describes only hopping processes within the LHB. In the LHB only
an electron at nearest neighbor site of the hole can hop by
interchanging its position with the hole.\cite{Dag08} The
remaining hopping processes describe either excitations to the
UHBs that are included in the superexchange ${\cal H}_J$ between
two V$^{3+}$ ions, or low spin charge excitations at the hole site
that contribute to the superexchange ${\cal H}_I$ for
V$^{4+}$-V$^{3+}$ pairs of ions, see below.

Further restriction on the hopping is introduced by the breaking of 
cubic symmetry in doped Y$_{1-x}$Ca$_x$VO$_3$ by the actual anisotropic
electron distribution over the $t_{2g}$ orbitals given by Eqs.
(\ref{frozen}). This leads to the symmetry breaking between the bonds
in $ab$ planes and along the $c$ axis.\cite{Kha04,Ole07} As we have
discussed in Sec. \ref{sec:Def}, hole doping occurs in the orbital
doublet $\{a,b\}$, and the $c$ orbitals are filled also in doped
systems by one electron each. Therefore, the $c$ electrons are
immobile in the strongly correlated regime and contribute only to
virtual $d^2_id^2_j\rightleftharpoons d^3_id^1_j$ excitations
which generate the superexchange along the considered bond
$\langle ij\rangle\parallel ab$.

In the large $U$ regime ($U\gg t$) the kinetic energy of the $\{a,b\}$
electrons is finite only in a doped system when the hopping process may 
occur in the restricted space. Furthermore, the hopping elements depend 
on the electronic configuration in $c$ orbitals. In case of empty $c$ 
orbitals, as e.g. in Sr$_2$VO$_2$, the hopping elements in Eq. 
(\ref{ht}) would be given by the bare tight binding element,
\cite{Dag08} i.e., $t_{\alpha}\equiv t$. On the contrary, the present 
situation with filled $c$ orbitals in doped Y$_{1-x}$Ca$_x$VO$_3$ 
perovskites resembles the case of doped manganites,
\cite{Dag01,Feh04,Tok06,Hor99,vdB99,Ali01,Ole02,Dag06,Sen06,Sti08}
where the hopping elements between active $e_g$ orbitals are
strongly renormalized by the $t_{2g}$ spins $S=3/2$ on both sites.
In the manganites this follows from strong Hund's exchange coupling 
$J_H$ between $e_g$ and $t_{2g}$ electrons which stabilizes a HS $S=2$ 
state at each Mn$^{3+}$ site.\cite{Fei99} Here one has instead a spin 
$S=1/2$ of a $c$ electron at each site which couples by Hund's exchange 
to the spin $s=1/2$ of a second electron in the $\{a,b\}$ doublet, and 
a HS state with spin $S=1$ arises. Therefore the constrainted hopping 
Hamiltonian $\tilde{\cal H}_t$ which
follows from the symmetry of $t_{2g}$ orbital states,
\begin{eqnarray}
\label{Ht}
\tilde{\cal H}_{t}=&-& t\sum_{\langle ij\rangle\parallel c}
({a}^{\dagger}_{i\sigma}{a}^{}_{j\sigma}
+{b}^{\dagger}_{i\sigma}{b}^{}_{j\sigma}+\mbox{H.c.})\nonumber \\
&-&t\sum_{\langle ij\rangle\parallel a}
({b}^{\dagger}_{i\sigma}{b}^{}_{j\sigma}+\mbox{H.c.})
 - t\sum_{\langle ij\rangle\parallel b}
({a}^{\dagger}_{i\sigma}{a}^{}_{j\sigma}+\mbox{H.c.})\,.\nonumber \\
\end{eqnarray}
and contains the operators which act in the projected space, with
electron number per site being either $n_i=2$ or $n_i=1$. When two
electrons are present ($n_i=2$), they interact by the local
exchange term $\tilde{\cal H}_{\rm Hund}$ in Eq. (\ref{tJ}), which
reads:
\begin{eqnarray}
\label{HJH}
\tilde{\cal H}_{\rm Hund}=-2J_H
\sum_{i}(\vec{S}_{ia}+\vec{S}_{ib})\cdot\vec{S}_{ic}\,.
\end{eqnarray}

Altogether, Eqs. (\ref{Ht}) and (\ref{HJH}) define the double
exchange model\cite{deG60} for strongly correlated $\{a,b\}$
electrons interacting with localized spins s=1/2 of $c$ electrons.
In the effective model which follows from it and is described in
Sec. \ref{sec:hole}, the effective hopping amplitude $t_{ij}\le t$
for the mobile $\{a,b\}$ electron does depend on the directions of
two electron spins in $c$ orbitals along a bond $\langle
ij\rangle$. We analyze the kinetic energy obtained in both AF
phases below in Sec. \ref{sec:hole}.

\subsection{Orbital-charge interaction near Ca defects }
\label{sec:opa}

\begin{figure}[t!]
\includegraphics[width=8.2cm]{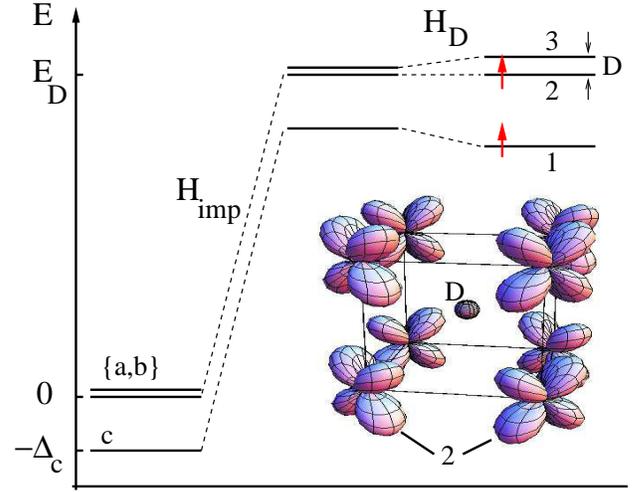}
\caption{(Color online) Energy levels of the $t_{2g}$ orbital
states $\gamma=a,b,c$ for a representative V$^{3+}$ ion in YVO$_3$
(left), and their modification in the vicinity of a Ca defect by
the impurity potential $H_{\rm imp}$ (middle) and after adding the
charge-orbital interaction $H_D$ (right). Two orbitals $\alpha=1$
and 2 are occupied at V$^{3+}$ ions in the vicinity of the defect
site $D$, as indicated by electron spins (arrows) which contribute
to spin $S=1$. Occupied $t_{2g}$ orbitals $\alpha=2$ with
intermediate energy are shown in the inset; there orbitals are
linear combinations of $\{|a\rangle,|b\rangle\}$ orbitals
$|\pm\rangle_i$, defined in Eq. (\ref{pm}). Orbital phases and the
$c$ orbitals occupied at each site are not shown for clarity. In
the doped case one of the occupied orbitals $\alpha=2$ is replaced
by a hole in the vicinity of the defect $D$ (see Fig. 1). }
\label{fig:imp}
\end{figure}

When an Y ion in  Y$_{1-x}$Ca$_x$VO$_3$ is replaced by a Ca impurity, 
the lattice is disturbed and two interaction terms arise due to the 
presence of the impurity. The first of them is the Coulomb potential 
due to the Ca impurity, introduced before in Eq. (\ref{Himp}), while 
the second one is a crystal field term ${\cal H}_D$ considered here 
as the second last term in the spin-orbital $t$-$J$ Hamiltonian Eq. 
(\ref{tJ}). The former term causes that a hole
in the VO$_3$ subsystem is confined to the immediate neighborhood
of the Ca$^{2+}$ charge defect in the Y$^{3+}$ sublattice by the
electrostatic potential Eq. (\ref{Himp}), as we have verified by
numerical calculations reported in Sec. \ref{sec:Def}. The latter
term originates from the quadrupolar component of electrostatic
field generated by a charge defect at the V$^{3+}$ ions. Replacing
an Y$^{3+}$ ion by a Ca$^{2+}$ ion implies that an effective {\it
negative charge\/} $e$ is introduced at the Ca impurity, i.e., in
the center of the cube shown in the inset of Fig. \ref{fig:imp}.
It interacts with an electron in the $\{a,b\}$ orbital doublet of
the considered V$^{3+}$ ion, and the repulsive energy between this
ion and the Ca defect has to be minimized. This may be achieved by
an optimal choice of the occupied $t_{2g}$ orbital in the
$\{a,b\}$ subspace.

Thus, the repulsive Coulomb potential of the Ca defect generates
an orbital polarization at surrounding it V ions with a pronounced
tendency toward electron occupation of one of the two linear
combinations of the active $\{a,b\}$ orbitals at site $i$,
\begin{equation}
\label{pm}
|\pm\rangle_i\equiv\frac12 \left(a_i^{\dagger}\pm
b_i^{\dagger}\right)|0\rangle\,,
\end{equation}
being eigenstates of $\sigma^x_i$ Pauli matrix. A properly chosen
linear combination, either $|+\rangle_i$ or $|-\rangle_i$ state at
site $i$, maximizes the average distance between the interacting
electronic charges at the V$^{3+}$ ion and at the Ca$^{2+}$
impurity, and minimizes the electrostatic interaction energy. 
The orbital states which satisfy this condition and are favored at
V$^{3+}$ ions are shown in the inset of Fig. \ref{fig:imp}. The
lifting of orbital degeneracy will be described by the following
orbital-defect polarization term acting in the $\{a,b\}$
sector,\cite{orb2}
\begin{equation}
\label{HD} {\cal H}_D=D\sum_{i\in {\cal C}_1}\lambda_i^{}\tau_i^x\,,
\end{equation}
where $\tau_i^x$ is defined in Eqs. (\ref{Pauli}). The summation
over $i\in {\cal C}_1$ in Eq. (\ref{HD}) includes the V sites which
belong to the cube ${\cal C}_1$ around the considered Ca site, and
the sign factor $\lambda_i=\pm 1$ (for $D>0$) selects the proper
orbital state which minimizes the charge-orbital interaction at
each site $i$. This polarized orbital is labeled as $\alpha=2$ in 
Fig. \ref{fig:imp}. It is occupied in V$^{3+}$ ions, but an extra 
hole introduced by doping does remove the electron from this topmost 
occupied orbital at a V$^{3+}$ ion.

As an illustrative example of the expected consequences of the 
orbital-defect polarization interaction $D$, see Eq. (\ref{HD}), we 
present the change of the orbital state for a single bond in Appendix 
\ref{sec:bond}. One finds a first order transition from the orbital
singlet with fluctuating orbitals to a fully polarized orbital state
at large $D$. We show below that although the considered bond interacts 
with its neighbors and orbital fluctuations are reduced from those in 
a single bond, the main features of this orbital transition survive 
also in a crystal.

\subsection{Superexchange for V$^{4+}$-V$^{3+}$ bonds}
\label{sec:sho}

As already discussed in Sec. \ref{sec:rel}, doping by a hole 
transfers a V$^{3+}$ ion into a V$^{4+}$ ion and excitations to 
the UHB are then removed at the hole site, see Fig.
\ref{fig:weight}. For the orbital background this implies a charge
defect and one may consider the problem of hole propagation,
either in the orbital model,\cite{Dag08} or in the spin-orbital
strong-coupling model.\cite{Woh09} Simultaneously, however,
superexchange interactions change in a drastic way and may be
treated in an analogous way as the superexchange between Mn$^{3+}$
and Mn$^{4+}$ ions in doped manganites.\cite{Ole02} Here the
superexchange couples spin $S=1$ of a V$^{3+}$ ion with spin
$s=1/2$ at the hole site, and may be derived by realizing that:
(i) only charge excitations $d_i^1d_j^2\rightleftharpoons
d_i^2d_j^1$ along the bond contribute and they depend on the
orbital degree of freedom, (ii) only charge excitations to the LS
($S=0$) states contribute and can be treated in perturbation
theory, and (iii) the excitation energies are given by Hund's
exchange $J_H$ as in the manganites.\cite{Ole02} There are two
$d^2$ singlet states: (i) interorbital singlets with energy
$(U-J_H)$, excited in $c_i^1c_j^1(a/b)_j^1\rightleftharpoons
c_i^1(a/b)_i^1c_j^1$ processes,\cite{noteab} and (ii) double
occupancies of the $c$ orbitals, excited in
$c_i^1c_j^1(a/b)_j^1\rightleftharpoons c_i^2(a/b)_j^1$ processes.
Due to the symmetry of the Coulomb interactions in the $t_{2g}$
subspace, the latter excitations contribute via two eigenstates
with energies $(U-J_H)$ and $(U+2J_H)$, respectively.\cite{Ole05}
Therefore, taking the energy difference with respect to the energy
$(U-3J_H)$ of the initial HS state at site $j$, one finds two
excitation energies: $2J_H$ and $5J_H$.

Note that for the particular occupancy $c^1(a/b)^1$ of $t_{2g}$
orbitals realized for V$^{3+}$ ions in YVO$_3$, see Eq.
(\ref{abc}), the $c_i^2$ excited states are generated only in
charge excitations along the bonds in the $ab$ plane, while the
former interorbital $c_i^1(a/b)_i^1$ singlets may arise on the
bonds along all three directions. Therefore, the superexchange
between the hole site (V$^{4+}$ ion) and the neighboring undoped
(V$^{3+}$) sites is anisotropic. More details concerning the
derivation of the superexchange between the $s=1/2$ spin of
V$^{4+}$ ion and its V$^{3+}$ neighbors with $S=1$ spins are
presented in Appendix \ref{sec:se01}.

While the superexchange in the $ab$ plane depends in principle on
the orbital $\{a,b\}$ flavor at the $d^2$ site and on the bond
direction, i.e., whether or not the hopping of the electron with
either $a$ or $b$ orbital flavor is allowed (or not) along the 
considered bond $\langle ij\rangle$, see Appendix \ref{sec:se01}, 
we present here the superexchange after averaging over the orbital
configuration at the $d^2$ site. This simplification is well
justified as we consider below (in Sec. \ref{sec:hole}) the
effective 1D models along the $c$ axis, where the spin and orbital
configurations at the neighboring sites along the bonds parallel
to either $a$ or $b$ axes are averaged out. Therefore, we
introduce the exchange constants:
\begin{equation}
\label{ex10} I_c \equiv \frac{t^2}{4J_H}\,,\hskip .7cm
I_{ab}\equiv \left(\frac{t^2}{8J_H}+ \frac{2t^2}{5J_H}\right)\,,
\end{equation}
and write the superexchange terms as follows,
\begin{eqnarray}
\label{HI}
{\cal H}_I &=& I_c\sum_{\langle i,j\rangle\parallel c}
\left({\vec s}_i\cdot {\vec S}_{j}-\frac12\right)
(1-n_i)n_j \nonumber\\
&+&I_{ab}\sum_{\langle ij\rangle\parallel ab}
\left({\vec s}_i\cdot {\vec S}_j-\frac12\right)(1-n_i)n_j\,.
\end{eqnarray}
Here the spin operators ${\vec s}_i$ refer to $s=1/2$ spin at the
hole site), while ${\vec S}_j$ stands for $S=1$ spins of the
undoped V$^{3+}$ sites neighboring with the hole site, so the
superexchange energy contributes for the bonds between V$^{3+}$
and V$^{4+}$ ions, i.e., when $(1-n_i)n_j\equiv 1$. This result is
used below (in Sec. \ref{sec:hole}) to derive and investigate 1D
orbital models which contain one doped hole.

\section{ One-dimensional orbital physics }
\label{sec:orbi}

\subsection{ Motivation and calculation method }
\label{sec:motiv}

In this section we analyze the perturbation of the spin-orbital 
structure due to the presence of defects, both for the $G$-AF and
the $C$-AF phase.
The main difference between the two phases arises from the magnetic
order along the $c$ axis, being either AF (FM) 
in the $G$-AF ($C$-AF) phase of Y$_{1-x}$Ca$_x$VO$_3$. On the contrary,
in the $ab$ plane both phases share a common AF and AO structure.
Thus for our estimate of the difference of the free energies the
orbital correlations along the $c$ axis will be of particular
importance.

The microscopic model Hamiltonian given in Eq. (\ref{tJ}) is too
complex to treat simultaneously spin and orbital dynamics in a
controlled approximation. Previous work has shown that larger
$S=1$ spins have weaker quantum effects and could be treated
classically, in contrast to the orbital $\tau=1/2$ pseudospins,
which undergo strong orbital quantum fluctuations in the $C$-AF phase
along the $c$ axis, and play a crucial role to explain its
anisotropic magnetic properties, \cite{Kha01,Ole07} the temperature 
dependence of the optical spectral weights,\cite{Kha04} and the phase 
diagram of the $R$VO$_3$ perovskites.\cite{Hor08} We emphasize that 
the bonds in $ab$ planes are AF and thus similar in both magnetic 
phases. To capture the difference between the $G$-AF and $C$-AF phase 
we focus here on the 1D orbital models along the $c$ axis derived from 
the microscopic spin-orbital model Eq. (\ref{tJ}). These orbital
models are solved below for finite 1D clusters coupled to the mean 
field (MF) terms which arise from the bonds in $ab$ planes and capture 
the main difference between both magnetic phases, the $G$-AF and $C$-AF 
phase of Y$_{1-x}$Ca$_x$VO$_3$. We analyze these models below in Secs.
\ref{sec:orbi} and \ref{sec:hole}. Note that this approach allows one 
to include the leading quantum fluctuations in the ground state; a 
similar cluster calculation was used recently to investigate the phase
diagram of the Kugel-Khomskii model for a bilayer.\cite{Brz11}

The orbital chains derived from the full spin-orbital model Eq.
(\ref{tJ}) are solved self-consistently using the MF
terms arising from the interactions with the neighboring V$^{3+}$
ions along the bonds parallel to $a$ and $b$ axes. This
simplification arises when bond orbital correlations are treated
classically and replaced by their MF values. The orbital order
parameter is then defined as
\begin{equation}
\label{OOP}
\langle\tau^z\rangle_{X}=\frac{1}{N_0}\sum_{\bf i}\langle\tau^z_{\bf i}\rangle
 e^{i \vec{Q}_{X}\cdot\vec{r}_{\bf i}}\,,
\end{equation}
where $X=C,G$ and $\vec{Q}_{C}=(0,\pi,\pi)$,
$\vec{Q}_{G}=(\pi,\pi,\pi)$ are vectors from the reciprocal space that
correspond to the orbital alternation in the $C$-AO and $G$-AO phase,
respectively. This approach is well justified here as the interactions
in $ab$ planes are Ising-like. Furthermore, it allows one to focus on
the quantum fluctuations along the 1D orbital chains parallel to the $c$
axis, where both $\{a,b\}$ orbital flavors are active, and on the role
played by the orbital polarization term Eq. (\ref{HD}).

\begin{figure}[t!]
\includegraphics[width=7.0cm]{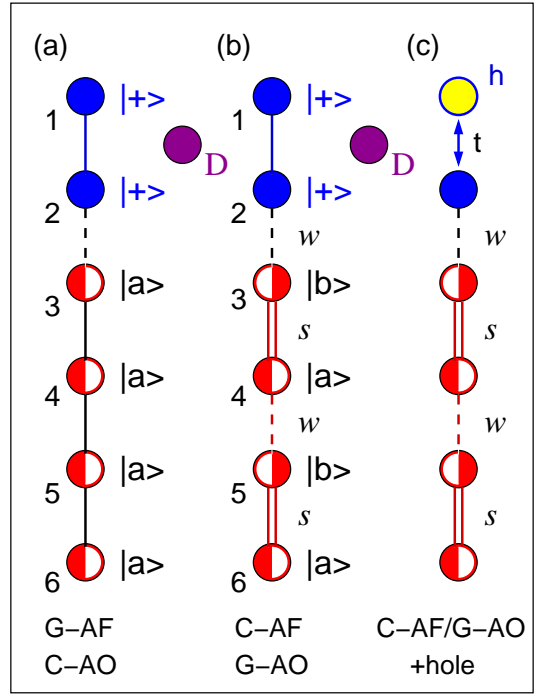}
\caption{(Color online)
Schematic view of a section of an orbital chain parallel to the $c$ 
axis. A Ca defect $D$ in the vicinity of the chain is shown in the top 
part, close to the bond $\langle 12\rangle$, and marked by dark filled 
circles; more distant nonequivalent sites are labeled as 
$i=3,\cdots,6$. The orbital chains correspond to: 
(a) the $G$-AF phase with $a$ electrons in $|a\rangle$ states 
(dark semicircles) corresponding to $C$-AO order along the chain 
(here sites $i=1$ and $i=2$ are equivalent); 
(b) the $C$-AF phase with $G$-AO order represented by alternating 
$\{|a\rangle,|b\rangle\}$ occupied orbital states along the chain 
(dark semicircles); 
(c) the $C$-AF phase with $G$-AO order of $\{a,b\}$ orbitals as in (b) 
and a hole $h$ at site $i=1$ near the defect $D$. 
The orbitals on two top sites
$i=1,2$ belong to the V$_8$ cube around Ca defect $D$ (see Fig.
\ref{fig:imp}), and are modified toward the $|+\rangle$ orbital
states by the increasing orbital polarization interaction $D$, see
Eq. (\ref{HD}). The hole in (c) is confined to sites $i=1,2$ by
the trapping Coulomb potential $E_D$. Away from the defect $D$,
orbital correlations in (a) are (almost) uniform and support AF
spin coupling, while in (b) and (c) they generate instead FM
interactions that alternate between strong ($s$) and weak ($w$)
exchange bonds. } \label{fig:orbi}
\end{figure}

\subsection{ Orbital chain in the $G$-AF phase }
\label{sec:gaf}

We consider first the undoped $G$-AF phase with classical spin order,
\begin{equation}
\label{ssg} \langle{\vec S}_i\cdot{\vec S}_{i+1}\rangle_c=-1,
\hskip .7cm \langle{\vec S}_i\cdot{\vec S}_j\rangle_{ab}=-1\,,
\end{equation}
and $C$-AO order, stable at low temperature in undoped YVO$_3$.
Orbital interactions in the $ab$ planes will be included in form of
MF terms which contain contributions both from the superexchange
$\propto J$ and from the Jahn-Teller-type orbital interactions
$\propto V_{ab}$. These two interactions support each other,
similar to the situation encountered in LaMnO$_3$.\cite{Fei99} For
the present $G$-AF phase with $C$-AO order the MF terms acting on
every site of an orbital chain along the $c$ axis are the same and
have alternating sign between two neighboring chains. We consider
here a representative orbital chain with the $C$-AO order 
with the orbital order parameter,
$\langle\tau^z\rangle_C>0$ defined in Eq. (\ref{OOP}) and
stabilized by an effective field, 
\begin{equation}
\label{hC}
h_C=\{2J\eta(r_1+r_3)+4V_{ab}\}\langle\tau^z\rangle_C\,.
\end{equation}
The field originates from four bonds perpendicular to the chain and 
belonging to an $ab$ plane and acts on the orbital pseudospins 
$\{\tau^z_i\}$ within the chain, see below. The effective 1D orbital 
model for the orbital chain along the $c$ axis within the 
$G$-AF phase takes therefore the form,
\begin{eqnarray}
\label{gaf1D}
{\cal H}_{G}^{1D}&=&\frac{1}{6}\,J\left(2r_1+1-3r_3\right)\sum_{i=1}^N
\vec\tau_i\cdot\vec\tau_{i+1}     \nonumber \\
&-&V_c\sum_{i=1}^N\tau^z_i\tau^z_{i+1} - D\sum_{i=1}^2\tau_i^x
-h_C\sum_{i=1}^N\tau_i^z\,.
\end{eqnarray}
The sign selected in the last term $\propto h_C$ gives indeed a
positive orbital order parameter $\langle\tau^z\rangle_C$ Eq.
(\ref{OOP}) when calculated self-consistently for the considered
orbital chain. The charge-orbital interactions $\propto D$ acts only
at the bond $\langle 12\rangle$ which belongs to a cube surrounding a
Ca site, i.e., at sites $i=1$ and $i=2$, and favors the orbitals
shown in Fig. \ref{fig:imp}. A schematic view of the orbital chain
in the $G$-AF phase is presented in Fig. \ref{fig:orbi}(a).

\begin{figure}[t!]
\includegraphics[width=7.2cm]{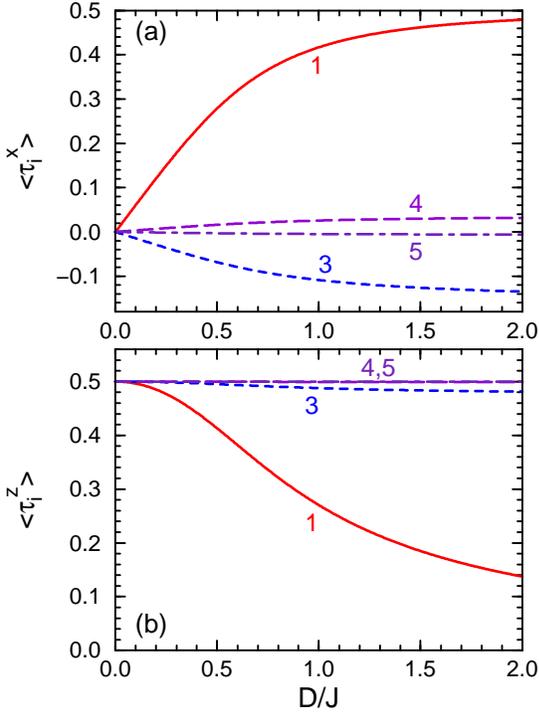}
\caption{(Color online)
On-site orbital expectation values for the $G$-AF phase for increasing
charge-orbital interaction $D$, see Eq. (\ref{HD}):
(a) orbital polarization $\langle\tau_i^x\rangle$, and
(b) orbital order parameter $\langle\tau_i^z\rangle$.
Labels of different curves indicate nonequivalent sites $i=1,3,4,5$
in a chain of $L_z=8$ sites, see Fig. \ref{fig:orbi}(a).
Parameters: $V_{ab}=0.2J$, $V_c=0.7J$.
}
\label{fig:xzg}
\end{figure}

We emphasize that only finite orbital interaction $V_c>0$
accompanied by the planar field $\propto h_C>0$ stabilizes the
observed $C$-AO order in the $G$-AF phase, while the superexchange
term alone (at $V_c=0$ and $h_C=0$) would favor instead alternation 
of occupied $\{a,b\}$ orbitals for finite Hund's exchange $\eta>0$,
with $\langle\vec\tau_i\cdot\vec\tau_{i+1}\rangle<0$. For the
parameters of Fig. \ref{fig:xzg} one finds that orbital fluctuations 
along the orbital chain are quenched in the $C$-AO order and the FO 
order is classical with $\langle\tau^z\rangle_C=0.5$ in absence of 
charge-orbital interaction. This result is independent of the chain 
length $L_z$. We show below how the classical $C$-AO order is 
modified by finite charge-orbital interaction $D$ which we treat 
here as a free parameter.

When a charge defect is introduced, the FO order in the chain is
locally disturbed, and the occupied orbitals on the two equivalent 
sites $i=1$ and $i=2$ of the bond $\langle 12\rangle$ gradually rotate 
with increasing $D$ toward the linear combinations which minimize the 
charge-orbital interaction, see the inset of Fig. \ref{fig:imp}. 
One thus expects that a final state obtained for sufficiently large 
polarization interaction $D$ has fully polarized orbitals, as shown
schematically in Fig. \ref{fig:orbi}(a). The FO order along the orbital
chain parallel to the $c$ axis remains undisturbed away from the Ca
defect, while close to it the occupied orbitals change to $|+\rangle$
orbitals for the sign selected in Eq. (\ref{gaf1D}).

The above scenario was confirmed by an exact diagonalization of
orbital chains of size $L_z=8$ sites which demonstrated local nature of
the perturbation introduced by a Ca impurity. The orbital state near 
the charge defect changes gradually under increasing interaction $D$, 
as shown in Fig. \ref{fig:xzg}. (We have verified that the length of 
$L_z=8$ sites is sufficient and longer chains of $L_z=10$ or 12 sites 
give practically the same numerical results.) For the present case
of classical $C$-AO order, the bond $\langle 12\rangle$ decouples
easily from the chain and the occupied orbitals at sites $i=1,2$ are
rotated, as shown by increasing expectation value of
$\langle\tau_1^x\rangle$, see Fig. \ref{fig:xzg}(a). This local
change near the defect is accompanied by weak negative polarization at
the next nearest neighbors of the defect (site $i=3$), and has
practically no effect at more distant sites $i=4$ and $i=5$.
The orbital moments near the defect,
$\langle\tau_1^z\rangle=\langle\tau_2^z\rangle$, are gradually
suppressed only near the defect, while the $\langle\tau_i^z\rangle$ ones
remain close to 0.5 away from it (for $i>2$)
[Fig. \ref{fig:xzg}(b)]. Therefore, the AF spin interactions along the
chain, supported by FO order, are only weakened between the first and
second neighbor of the defect site, i.e., at the bond
$\langle 23\rangle$ in Fig. \ref{fig:orbi}(a). Altogether, the
numerical results demonstrate that the orbital state is disturbed only
locally near a Ca defect and these local modifications of the orbital 
state do not destabilize the coexisting $G$-AF and $C$-AO order away 
from the defect.

\subsection{ Orbital chain in the $C$-AF phase }
\label{sec:caf}

In order to derive the form of the orbital chain for the $C$-AF phase
we use the following spin correlations in the classical state:
\begin{equation}
\label{ssc} \langle{\vec S}_i\cdot{\vec S}_{i+1}\rangle_c=1\,,
\hskip .7cm \langle{\vec S}_i\cdot{\vec S}_j\rangle_{ab}=-1\,.
\end{equation}
The FM spin correlation function along the $c$ axis suppresses
then all low-spin contributions to the superexchange Eq.
(\ref{HJ}) and the orbitals participate in strong quantum
fluctuations along the chain, induced by the SU(2)-symmetric
interaction $\propto\vec\tau_i\cdot\vec\tau_{i+1}$ with a large
prefactor $Jr_1$. The model Hamiltonian which describes the 1D
orbital chain in the $C$-AF phase shown in Fig. \ref{fig:orbi}(b)
takes a similar form to the one for the $G$-AF phase discussed in
Sec. \ref{sec:gaf}:
\begin{eqnarray}
\label{caf1D}
{\cal H}_{C}^{1D}&=&Jr_1\sum_{i=1}^N\vec\tau_i\cdot\vec\tau_{i+1}
-V_c\sum_{\langle i,i+1\rangle\parallel c}\tau^z_i\tau^z_{i+1}\nonumber \\
&-& D\sum_{i=1}^2\tau_i^x-h_G\sum_{i=1}^N(-1)^i\tau_i^z\,.
\end{eqnarray}
The $G$-AO order is here stabilized again by the planar field due to
the vanadium neighbors in the $ab$ plane:
\begin{equation}
\label{hG}
h_G=\{2J\eta(r_1+r_3)+4V_{ab}\}\langle\tau^z\rangle_G\,,
\end{equation}
which is here proportional to the orbital order parameter
$\langle\tau^z\rangle_G$, alternating along the chain, see Eq.
(\ref{OOP}). However, in contrast to the classical FO order considered
in Sec. \ref{sec:gaf}, the orbital order is here rather weak as it
competes with the orbital fluctuations along the chain.\cite{Kha04}
Thus the ground state is stabilized by strong $\{a,b\}$ quantum orbital
fluctuations, while the real orbital order parameter
$\langle\tau^z\rangle_G$ is reduced. Moreover, the orbital interactions
$\propto V_c$ along the chain favor $C$-AO order and are in conflict
with the $G$-AO order in the present case. This reduces the orbital
order parameter $\langle\tau^z\rangle_G$ further as the field $h_G$ is
considerably smaller than $h_C$, see Eq. (\ref{hC}).

\begin{figure}[t!]
\includegraphics[width=7.2cm]{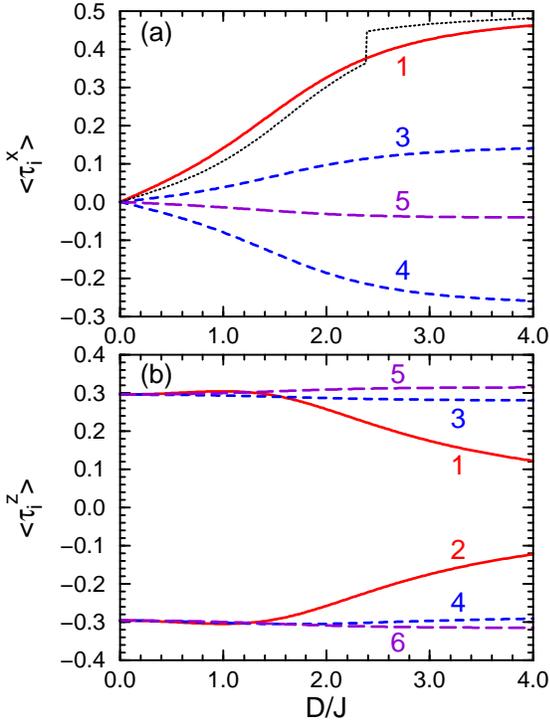}
\caption{(Color online)
On-site orbital expectation values obtained for the $C$-AF phase as a
function of increasing charge-orbital interaction $D$, see Eq.
(\ref{HD}), obtained for a chain of $L_z=8$ sites:
(a) orbital polarization $\langle\tau_i^x\rangle$, and
(b) orbital order parameter $\langle\tau_i^z\rangle$.
Dotted line in (a) shows orbital polarization $\langle\tau_i^x\rangle$
for the free chain (at $h_{ab}=0$).
Labels of different curves and parameters as in Fig.~\ref{fig:xzg}.
}
\label{fig:xzc}
\end{figure}

First, we performed a self-consistent calculation for a reference chain 
of $L_z=8$ sites in absence of Ca defects, i.e., with no polarization 
term ($D=0$), which gives the orbital order parameter
$\langle\tau^z\rangle_G\simeq 0.296$ for the present parameters,
see Fig. \ref{fig:xzc}(b).
Next we considered the orbital chain given by Eq. (\ref{caf1D}) near
the charge defect, with orbitals polarized near the defect ($D>0$) and
coupled to the neighboring sites in $ab$ planes by the MF terms.
The relevant $G$-AO order is shown in Fig. \ref{fig:orbi}(b). Due to
the strong $\{a,b\}$ orbital fluctuations along the chain, the orbital
state changes here much slower with increasing charge-orbital
interaction $D$ than in the $G$-AF phase, see Fig. \ref{fig:xzc}.
The increasing orbital moments
$\langle\tau^x_1\rangle=\langle\tau^x_2\rangle$ near the charge
defect gradually disturb the fluctuations along the chain and induce
also finite $\langle\tau^x_i\rangle$ moments on more distant sites
which alternate along the chain, see Fig. \ref{fig:xzc}(a). In contrast,
the $\langle\tau^z_i\rangle$ moments are first undisturbed as long
as the change in the orbital ground state is small, but next they
are gradually suppressed at the nearest neighbor sites of the charge
defect when $D$ increases beyond $D>J$, cf. Appendix \ref{sec:bond}. 
Away from the defect the weak orbital order is practically undisturbed 
due to the MF terms, while near the charge defect, i.e., on the bond 
$\langle 12\rangle$, the orbital
state is locally modified by the charge-orbital interaction for $D>2J$.

\begin{figure}[t!]
\includegraphics[width=7.2cm]{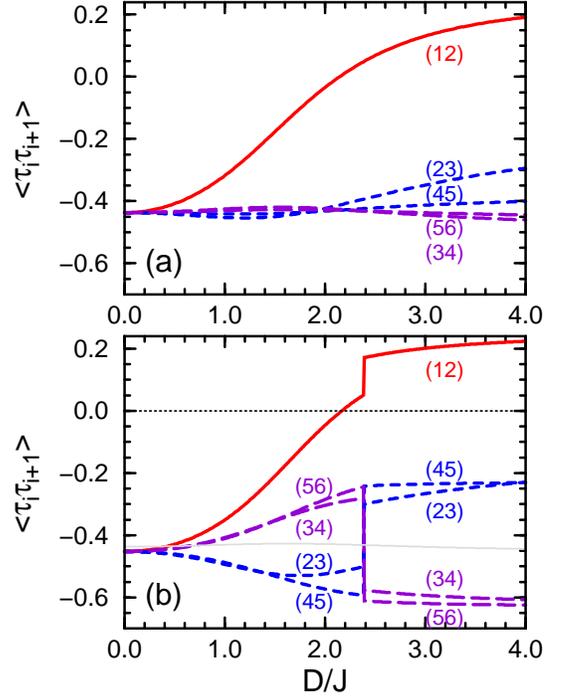}
\caption{(Color online)
Orbital correlation functions
$\langle\vec\tau_i\cdot\vec\tau_{i+1}\rangle$ found for the orbital
chain with $L_z=8$ along the $c$ axis in the $C$-AF phase shown in
Fig. \ref{fig:orbi}(b):
(a) self-consistent calculation with the MF term $\propto h_C$ Eq.
(\ref{hC}) acting on the chain from the neighboring sites in the $ab$
planes;
(b) free orbital chain ($h_C=0$).
Parameters: $V_{ab}=0.2J$, $V_c=0.7J$, $t=6.25J$.
}
\label{fig:tt1}
\end{figure}

Further evidence that the bond $\langle 12\rangle$ close to the charge
defect decouples from the orbital chain when the charge-orbital
interaction is sufficiently large is given by the orbital correlations
$\langle\vec\tau_i\cdot\vec\tau_{i+1}\rangle$. For the present
parameters one finds
$\langle\vec\tau_i\cdot\vec\tau_{i+1}\rangle\simeq-0.437$ for the
reference chain of $L_z=8$ sites in absence of charge defect ($D=0$),
see Fig. \ref{fig:tt1}(a). The orbital fluctuations are only weakly
reduced by the $G$-AO order from the value of
$\langle\vec\tau_i\cdot\vec\tau_{i+1}\rangle\simeq-0.452$ obtained
for a free chain, see Fig. \ref{fig:tt1}(b).

When charge-orbital interaction is introduced ($D>0$), the orbital
fluctuations near the defect, $\langle\vec\tau_1\cdot\vec\tau_2\rangle$,
are first reduced by the perturbative positive term $\propto D^2$ when
$D\ll J$, but soon (at $D\sim J$) their reduction becomes close to
linear in $D$, and finally this bond decouples from the orbital chain
(for $D\sim 2J$) as the orbital fluctuations are locally suppressed by
uniformly polarized orbitals, see Fig. \ref{fig:tt1}(a). Therefore, the
correlation function $\langle\vec\tau_1\cdot\vec\tau_2\rangle$ changes
sign at $D\simeq 2.15J$ and becomes positive for larger values of $D$,
approaching the limit of fully polarized orbitals, with FO correlation
near the defect, $\langle\vec\tau_1\cdot\vec\tau_2\rangle=0.25$. At
the same time, the remaining correlations are only weakly modified and
correspond approximately to those obtained for a shorter orbital
chain with open ends under the influence of the MF terms.
The largest change is found for the
$\langle\vec\tau_2\cdot\vec\tau_3\rangle$ correlation which
approaches the classical value $-0.25$ when $D>4J$ (not shown),
but the remaining orbital correlations are only weakly influenced
by the presence of the charge defect.

The essential feature of the evolution of orbital correlations
with increasing orbital-charge interaction $D$ is the gradual decoupling
of the bond $\langle 12\rangle$, next to the charge defect,
from the fluctuating orbital chain. This phenomenon is more pronounced
for a free orbital chain, shown in Fig. \ref{fig:tt1}(b). In this case
one finds a somewhat surprising effect of induced dimerized
correlations along the chain, as explained below.
In addition to the perturbative regime of weak $D<J$, where
the orbital correlations are modified by terms $\propto D^2$, one
recognizes here two distinct regimes, separated by a critical value
$D_c\simeq 2.39J$ at which the ground state changes abruptly and the
bond $\langle 12\rangle$ polarized by the charge-orbital interaction
decouples from
the fluctuating chain, see Fig. \ref{fig:orbi}(b). At $D=D_c$ a quantum
transition takes place from a jointly fluctuating orbital chain to
broken chain with its ends neighboring with a
static orbital-polarized bond near the charge defect.
We remark that this transition from fluctuating to polarized orbitals
near the charge defect is similar to the one which takes place for a
single bond, see Appendix \ref{sec:bond}. Common features are that the
transition is also discontinuous here and occurs to the state with the
orbitals near the charge defect being almost fully polarized, with
$\langle\vec\tau_1\cdot\vec\tau_2\rangle>0.17$.

The two regimes separated by the above transition are quite
distinct. For $D<D_c$ the charge-orbital interaction disturbs
orbital fluctuations along the chain as they become gradually
suppressed on the bond $\langle 12\rangle$, and triggers dimerized
orbital correlations. Orbital fluctuations are then enhanced on
the bonds neighboring with the disturbed bond, represented here by
the bond $\langle 23\rangle$, see Fig. \ref{fig:orbi}(b). This
perturbation generates alternation between weaker and stronger
orbital correlations along the chain. In contrast, for $D>D_c$ the
orbitals freeze in the polarized state on the $\langle 12\rangle$
bond, and thus the orbital fluctuations become restricted by the
constraint imposed by this bond at both ends of the remaining
chain. Thus the role of stronger and weaker fluctuating bonds
along the broken chain is now reversed, and the fluctuations on
the bond $\langle 23\rangle$ become weak.

One finds that the above quantum transition for a free chain
changes to a crossover for an embedded chain, see Fig.
\ref{fig:tt1}(a). Also in this latter case it leads to dimerized
orbital correlations along the orbital chain but this alternation
is much weaker than for a free chain. This might suggest that
charge-orbital interaction could be responsible for dimerization
along the orbital chains, and would imply dimerized FM
interactions in the $C$-AF phase. When spin interactions are
evaluated for the orbital correlations of Fig. \ref{fig:tt1}(b),
FM interactions are indeed weak on bonds $\langle 23\rangle$ and
$\langle 45\rangle$ (shown by dashed lines) and strong on bonds
$\langle 34\rangle$ and $\langle 56\rangle$ (shown by long-dashed
lines). However, the coupling with other orbital chains along the
bonds in $ab$ planes reduces considerably these dimerized
interactions, see Fig. \ref{fig:tt1}(a). Nevertheless, we
introduced average orbital correlation functions for strong and
weak bonds, $\langle{\vec\tau}_i\cdot{\vec\tau}_{i+1}\rangle^o_s$
and $\langle{\vec\tau}_i\cdot{\vec\tau}_{i+1}\rangle^o_w$ as shown
in Fig. \ref{fig:orbi}(b), and found that they differ from each
other. We shall use them below in Sec. \ref{sec:dim} and
demonstrate that the dimerization is indeed predicted by the
present $t$-$J$ orbital model Eq. (\ref{tJ}), yet its mechanism is
more subtle. Apart from the orbital polarization in the vicinity
of a defect which we have discussed here, the orbital order is
also disturbed by a hole introduced with each defect.

\section{ Doped hole in orbital chains }
\label{sec:hole}

\subsection{ Orbital $t$-$J$ model for the $G$-AF phase }
\label{sec:de}

We first analyze the local changes of spin and orbital correlations
introduced by a hole in the $G$-AF phase. There are two
distinct processes how the hole acts on the spin-orbital structure,
namely via: (i) superexchange between s=1/2 and S=1 sites, and
(ii) double exchange which involves hole motion in the LHB
(with spins in the HS configuration).
The double exchange hopping process, i.e.,
$(\frac{1}{2},1)\rightarrow (1,\frac{1}{2})$, is controlled by the
spin orientation of the involved spins of electrons in $xy$ orbitals.
One should note that two DE processes, one forward and one backward,
are indeed distinct from the superexchange process
$(\frac{1}{2},1)\rightarrow (0,\frac{1}{2})\rightarrow (\frac{1}{2},1)$
which involves an $S=0$ intermediate state with an excitation energy
$\propto J_H$. Interestingly this latter proportionality makes this
kind of superexchange larger than the conventional superexchange
$\propto J$, i.e., between $S=1$ spins, which results from virtual
excitations across the Mott-Hubbard gap  $\propto U$. The formal
aspects of this kind of exchange were discussed in Sec. \ref{sec:sho}.

\begin{figure}[t!]
\includegraphics[width=5.5cm]{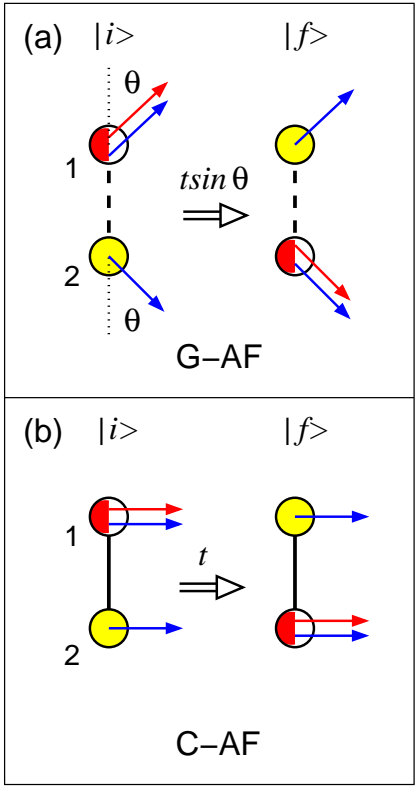}
\caption{(Color online) Schematic view of the hole hopping along
the bond $\langle 12\rangle$ belonging to the cube surrounding a
Ca defect for the: (a) $G$-AF phase, and (b) $C$-AF phase. In the
initial state $|i\rangle$ two spins $s=1/2$ due to electrons which 
occupy $|c\rangle$ and $|a\rangle$ orbitals form a high-spin $S=1$ 
state at $d^2$ site $i=1$. The $d^1$ site $i=2$ is occupied by the 
hole (filled circle) and has only a single $s=1/2$ spin due to the 
electron occupying the $|c\rangle$ orbital. In the $G$-AF phase shown 
in (a) a hopping process of the electron with $a$ orbital flavor, 
$\propto ta_{1}^{\dagger}a_{2}^{}$, generates a finite canting angle 
$\theta$ for both spins with respect to their original orientation
($\theta=0$) --- which leads to a finite kinetic energy gain
$\propto t\sin\theta$ by the transition to the final $|f\rangle$
high-spin $S=1$ state. In contrast, full hopping amplitude $t$
contributes along FM bonds in the $C$-AF phase, see (b). }
\label{fig:de}
\end{figure}

Due to large binding energy between the hole and  the Ca impurity,
the hole is confined to one of the topmost occupied orbitals of
the V$^{3+}$ cube shown in Fig. \ref{fig:imp}.
As discussed in Sec.\ref{sec:Def} the motion of the
the hole along the bond $\langle 12\rangle$ parallel to the $c$ axis
is of particular relevance.
A schematic picture is given in Fig. \ref{fig:orbi}(c), where one
of the polarized orbitals in Fig. \ref{fig:orbi}(a)
is replaced by a hole. The AF order in the $ab$ plane
is stabilized mainly by the superexchange driven by the excitations in
$c$ orbitals which are unaffected by the doped hole. This implies that
the hole motion is confined to the $\langle 12\rangle$ bond along the
$c$ axis, and the effective hopping element is determined here by the
double exchange.

As usual in the double exchange model,\cite{deG60} the AF order is
disturbed and spins cant in order to find a compromise between the loss
of the magnetic energy with respect to the AF spin order, and the
kinetic energy, which would be minimal for FM spins. In the present 
model with hole confinement near the charge defect, it suffices to 
analyze the double exchange mechanism on the bond $\langle 12\rangle$ 
along the $c$ axis, see Fig. \ref{fig:de}, supplemented by the magnetic 
energy contributions arising due to other bonds which start at either 
$i=1$ or $i=2$. In the initial state $|i\rangle$ shown in Fig. 
\ref{fig:de}(a) the spin of an $a\equiv yz$ electron at site $i=1$ is 
parallel to the one of the electron in $c$ orbital. The hopping process 
to the final state $|f\rangle$ with the $a$ electron moved to site $i=2$ 
is possible after the spins at both sites are canted by angle $\theta$ 
away from their AF order in the $G$-AF phase, see Fig. \ref{fig:de}(a). 
In this case the hopping amplitude is given by
\begin{equation}
\label{theta}
t(\theta)\equiv t\sin\theta\,.
\end{equation}

\begin{figure}[t!]
\includegraphics[width=7.2cm]{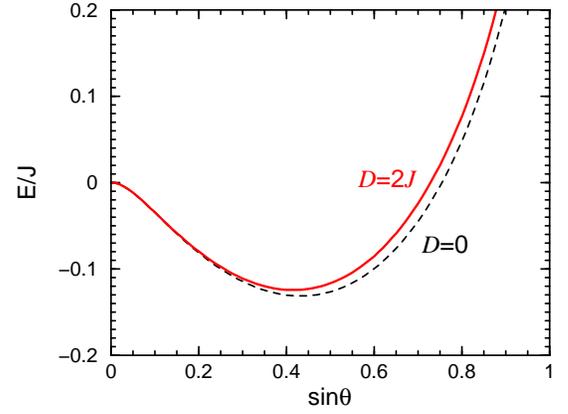}
\caption{(Color online)
Total energy per site ${\cal E}_G$ calculated for the $G$-AF state with 
$C$-AO orbital order as function of tilt angle $\sin\theta$, as 
obtained by exact diagonalization of an orbital chain of $L_z=8$ sites
with $D=2J$ (solid line) and $D=0$ (dashed line).
Parameters: $V_{ab}=0.2J$, $V_c=0.7J$, $t=6.25J$.
}
\label{fig:egaf}
\end{figure}

Consequently, the orbital chain containing a hole in the $G$-AF phase is
described by the orbital Hamiltonian (relevant for the $C$-AO order):
\begin{eqnarray}
\label{gafh}
{\cal H}_{G}^h(\theta)&=&- t(\theta)
({a}^{\dagger}_{1}{a}^{}_{2}
+{b}^{\dagger}_{1}{b}^{}_{2}+\mbox{H.c.})\nonumber\\
&+&Jr_1\sum_{i=1}^N \vec\tau_i\cdot\vec\tau_j
 - V_c\sum_{i=1}^N\tau^z_i\tau^z_j \nonumber \\
&-& D\sum_{i=1}^2\tau_i^x
-h_C\sum_{i=1}^N\tau_i^z\,.
\end{eqnarray}
As in Eq. (\ref{tJ}), the creation operators $\{a^\dagger,b^\dagger\}$
of spinless fermions act in the restricted space, and the hopping occurs
along the $\langle 12\rangle$ bond. The orbital chain depends on the
orbital MF $h_C$, see Eq. (\ref{hC}), and on the spin canting angle 
$\theta$ via the hopping $t(\theta)$, as given in Eq. (\ref{theta}). 

We performed exact diagonalization of the orbital chain Eq. (\ref{gafh})
for representative parameters of a doped Y$_{1-x}$Ca$_x$VO$_3$ system.
Thereby we investigated the total energy including the magnetic energy 
${\cal E}_G^{\rm mag}(\theta)$ which follows from the bonds which are 
influenced by the spin canting at sites $i=1$ and $i=2$, see Appendix C. 
As expected, one finds that the energy is lowered when the spin order
at the hole site and in its neighborhood is locally disturbed and
permits hopping with the reduced hopping element given by Eq.
(\ref{theta}). The kinetic energy $\propto t(\theta)$ is gained and 
part of the magnetic energy Eq. (\ref{emag}) is lost when the spins
cant and rotate away from their orientation in the ideal $G$-AF phase.
The optimal angle is found to be given by $\sin\theta\simeq 0.42$ at
$D=2J$ and $\sin\theta\simeq 0.43$ at $D=0$, see Fig. \ref{fig:egaf}.
Thus, the dependence of $\theta_0$ on the charge-orbital polarization
term is surprisingly weak and we may consider the angle
$\sin\theta_0\simeq 0.42$ obtained for $D=2J$ as a representative value
for the analysis of doped Y$_{1-x}$Ca$_x$VO$_3$, see Sec. \ref{sec:dim}.
We have found that already rather weak charge-orbital interaction $D<J$
polarizes almost entirely the occupied orbital within the 
$\langle 12\rangle$ bond
which decouples from the remaining part of the orbital chain.

It is worth noting that the spin structure is robust and its local
modification is moderate. This contradicts naive expectations based on 
rather small magnetic exchange constants between $S=1$ spins, being 
only a fraction of $J$ and determined by neutron experiments,
\cite{Ulr03} that the kinetic energy would dominate over magnetic one 
since $t\gg J$. Instead, we have found that the canting of the spin 
structure is only moderate for the realistic parameters, see Fig. 
\ref{fig:egaf}. This surprising result follows from large AF 
superexchange interactions in the vicinity of the hole, 
$I_c\simeq 1.92J$ and $I_{ab}\simeq 4.1J$, which are enhanced by small 
charge excitation energies $\propto J_H$. Here we used $J_H=0.64$ eV 
for $\eta=J_H/U=0.13$.\cite{Ole07} Indeed, these exchange constants
($I_c$ and $I_{ab}$) are larger by approximately one order of
magnitude than the exchange constants $J_c$ and $J_{ab}$ between $S=1$ 
spins.

\subsection{ Orbital $t$-$J$ model for the $C$-AF phase }
\label{sec:holec}

In contrast to the $G$-AF phase discussed before, the spin structure of
the $C$-AF phase is already tuned to optimize the double exchange, i.e.,
the hole hopping term is $t$ near the charge defect, see Fig.
\ref{fig:de}(b). An intriguing question here, however, is to what
extent the spin structure is affected indirectly via the perturbations
of the orbital chain. After introducing a hole, an orbital chain with an
even number of sites as considered in Sec. \ref{sec:caf} becomes an open
chain with an odd number of sites and an additional constraint
--- the orbital at one of its ends is polarized by the charge-orbital
interaction $\propto D$, and this orbital may interchange its
position with the hole.

\begin{figure}[t!]
\includegraphics[width=7.2cm]{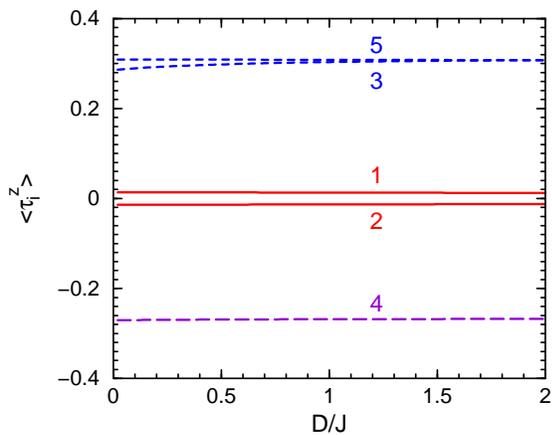}
\caption{(Color online)
Orbital order parameter $\langle\tau_i^z\rangle$ in the $C$-AF phase
around the charge defect neighboring with the bond $\langle 12\rangle$
shown in Fig. \ref{fig:orbi}(c), obtained for an embedded orbital
chain with finite field $h_G$ for increasing orbital polarization
interaction $D$, see Eq. (\ref{HD}).
Parameters: $V_{ab}=0.2J$, $V_c=0.7J$ and $t=6.25J$. }
\label{fig:tzh}
\end{figure}

The orbital Hamiltonian which describes a hole in the $C$-AF phase
(with $G$-AO order) is easily derived from Eq. (\ref{caf1D}),
\begin{eqnarray}
\label{cafh}
{\cal H}_{C}^h&=&- t ({a}^{\dagger}_{1}{a}^{}_{2}
+{b}^{\dagger}_{1}{b}^{}_{2}+\mbox{H.c.})   \nonumber \\
&+&Jr_1\sum_{\langle i,i+1\rangle\parallel c}
\vec\tau_i\cdot\vec\tau_{i+1}
 - V_c\sum_{\langle i,i+1\rangle\parallel c}\tau^z_i\tau^z_{i+1} \nonumber \\
&-& D\sum_{i=1,2}\tau_i^x-h_G\sum_i(-1)^i\tau_i^z\,.
\end{eqnarray}
It depends on the MF $h_G$ defined in Eq. (\ref{hG}) which stabilizes
the $G$-AO order and is relevant for the $C$-AF phase. Here we adopted
the usual notation that the bond $\langle 12\rangle$ belongs to the cube
surrounding the charge defect, see Fig. \ref{fig:orbi}(c).
As the lone electron at the undoped site is delocalized over the bond
$\langle 12\rangle$, the corresponding orbital is practically decoupled
from the orbital chain along the $c$ axis, and is easily polarized to
minimize the charge-orbital interaction by a rather weak interaction,
$D\ll J$. Therefore, the order parameter $\langle\tau_i\rangle$
vanishes at this bond, i.e., for $i=1$ and $i=2$, see Fig.
\ref{fig:tzh}. One finds as well that the orbital moments
$\langle\tau_i^z\rangle$ are undisturbed and alternate in the remaining
part of the chain, as expected for the $G$-AO order.

\begin{figure}[t!]
\includegraphics[width=7.2cm]{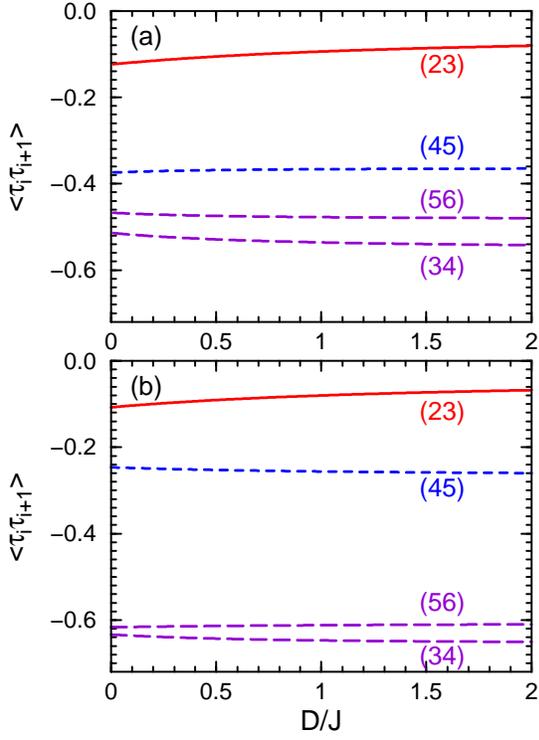}
\caption{(Color online) Alternation of the bond orbital correlations
$\langle\vec{\tau}_i\cdot\vec{\tau}_{i+1}\rangle$ as obtained for the
$C$-AF phase in the vicinity of the charge defect, see Fig.
\ref{fig:orbi}(c), as function of orbital polarization interaction $D$
Eq. (\ref{HD}).
The correlation functions which alternate between strong
($\langle 34\rangle$ and $\langle 56\rangle$, long-dashed lines) and
weak ($\langle 23\rangle$, solid lines, and $\langle 45\rangle$, dashed
lines) bonds, are obtained for an orbital chain with $L_z=8$ sites:
(a) embedded with finite $h_C$ ($V_{ab}=0.2J$), and
(b) free with $h_C=0$.
Other parameters: $V_c=0.7J$ and $t=6.25J$. }
\label{fig:ttdimh}
\end{figure}

The most important consequence of the fragmentation of the orbital
chain by the presence of a doped hole is the alternation of orbital
correlations $\langle{\vec\tau}_i\cdot{\vec\tau}_{i+1}\rangle$ shown in
Fig. \ref{fig:ttdimh}.
The hole due to the confinement by the defect potential moves
predominantly on the bond $\langle 12\rangle$ which leads to strong
orbital fluctuations on sites 1 and 2. These strong fluctuations
suppress the correlations on the neighbor bond $\langle 23\rangle$
(and on the equivalent bond $\langle 1 L_z\rangle$ under periodic
boundary conditions).

Therefore, bonds with enhanced and reduced orbital fluctuations
$|\langle{\vec\tau}_i\cdot{\vec\tau}_{i+1}\rangle|$ {\it
alternate\/} along the chain and, in analogy to Sec.
\ref{sec:caf}, we introduce the average orbital correlation
functions for strong and weak bonds:
$\langle{\vec\tau}_i\cdot{\vec\tau}_{i+1}\rangle_{hs}$ and
$\langle{\vec\tau}_i\cdot{\vec\tau}_{i+1}\rangle_{hw}$. This
alternation is most pronounced in a free chain, see Fig.
\ref{fig:ttdimh}(b), but is sufficiently robust to survive the
coupling of the chain to its neighboring sites in the $ab$ planes
of the $C$-AF phase, see Fig. \ref{fig:ttdimh}(a). Therefore, we
find here {\it dimerization\/} of the orbital interactions around
a hole which is the driving force toward dimerized FM spin
interactions in a doped $C$-AF phase, see Sec. \ref{sec:dim}. We
also note that the alternation of orbital bond strength induced by
the holes in the $C$-AF phase is reminiscent to the bond
alternation induced by large charge-orbital interaction $D$, see
Fig. \ref{fig:tt1}(b).

\section{ Scenario for the G-AF to C-AF magnetic transition }
\label{sec:num}

\subsection{ Statistical averaging in doped phases }
\label{sec:sta}

To demonstrate that a phase transition from the $G$-AF to $C$-AF phase
indeed occurs with increasing doping $x$ in the Y$_{1-x}$Ca$_x$VO$_3$
system, we use here a statistical approach designed for the
weakly doped regime. In the dilute limit we can neglect interactions
among defects. Thus it suffices to analyze configurations of orbital
chains of a fixed length with no more than one hole. A single calcium
charge defect which occurs in Y$_{1-x}$Ca$_x$VO$_3$ influences the
orbital order at eight vanadium ions which surround it in a cube shown
in the inset of Fig. \ref{fig:imp}.

We start with the $G$-AF phase with $C$-AO order.
This situation will be used as a reference state below and we shall
investigate deviations from it by introducing charge defects. All
orbitals in the orbital chains considered above are to some extent
influenced by the presence of the charge defect, as we have seen in the
previous Secs. \ref{sec:orbi} and \ref{sec:hole}. These modifications
occur in a different way, i.e., depending on whether the chain under
consideration is undoped (Sec. \ref{sec:orbi}) or contains a single
hole (Sec. \ref{sec:hole}). Among the direct neighbor chains of a defect,
one chain contains a hole and the remaining three chains are undoped.

Next we want to determine the probability to find a chain segment of
length $L_z$ that contains a hole. To this end we introduce the specific
volume of a defect $v_D=L^2L_z$. We assume here that the cubic lattice
of V ions has lattice constant $c=1$, such that the defect concentration
is $x=1/L^2L_z$, and the total number of $c$-chains threading the box is
$L^2$. As we are interested in the dilute limit here, we choose for the
box height $L_z=8$ which is also the length of chains studied by
numerical diagonalization. As $L$ is in the range $2,3,\cdots,L_z$, the
accessible concentration range is $x\in(0.002,0.03)$. Smaller values for
$L_z$ would allow us to study larger doping concentration $x$. This may
now be used to construct the statistical averaging over the bonds in a
doped system. We introduce statistical weights or probabilities,
\begin{equation}
\label{weight}
w_h\equiv \frac{1}{L^2}\,,\hskip .7cm w_e\equiv \frac{3}{L^2} \,,
\hskip .7cm w_0\equiv \frac{L^2-4}{L^2} \,,
\end{equation}
for finding a chain doped by a hole ($w_h$) or an undoped chain
next to a defect ($w_e$), respectively.
Chains separated from a cube with a defect site
occur with the complementary probability $w_0\equiv 1-w_h-w_e$.
These weights will be used below to:
(i) estimate to what extent dimerized interactions develop in the doped
$C$-AF phase (in Sec. \ref{sec:dim}),
(ii) obtain the energy of both AF phases which compete with each other,
and
(iii) investigate whether a phase transition could occur from the
$G$-AF to the $C$-AF phase in the low doping regime,
see Sec. \ref{sec:pht}.

\subsection{ Dimerization in orbital chains }
\label{sec:dim}

\begin{figure}[t!]
\includegraphics[width=7.2cm]{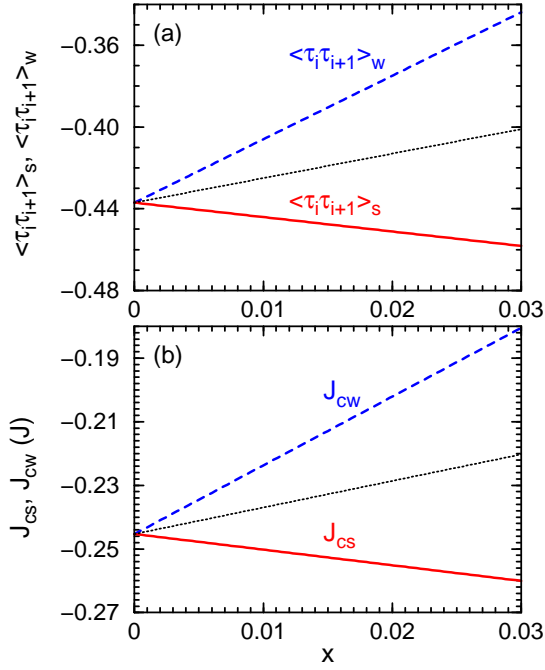}
\caption{(Color online)
Dimerization in the $C$-AF phase induced by charge defects as function
of the orbital polarization interaction $D$ Eq. (\ref{HD}):
(a) orbital correlations $\langle\vec{\tau}_i\cdot\vec{\tau}_{i+1}\rangle$
on {\it strong} ($i=3,5$, solid lines) and {\it weak bonds}
($i=2,4$, dashed lines), and
(b) magnetic exchange constants for {\it strong} ($J_{cs}$, solid line)
                           and {\it weak} ($J_{cw}$, dashed line)
FM bonds along the $c$ axis.
Dotted lines indicate the average values in both cases.
Parameters: $V_{ab}=0.2J$, $V_c=0.7J$, $t=6.25J$, $D=3J$.
}
\label{fig:ttdim}
\end{figure}

As we have shown in Sec. \ref{sec:holec}, a doped hole breaks the
orbital chains, suppresses locally the orbital fluctuations and thus
induces dimerized orbital correlations (see Fig. \ref{fig:ttdim}).
To some extent a similar dimerization occurs also in the undoped
orbital chains that are direct neighbors of charge defects (see Fig.
\ref{fig:tt1}). Although the dimerized orbital correlations occur in
the chain that contains a  hole, we assume that the magnetic order
and excitations will reflect an average alternation in the doped
Y$_{1-x}$Ca$_x$VO$_3$ crystal. Therefore, we average here the orbital
correlations which occur on stronger and weaker fluctuating orbital
bonds over the entire sample.

The orbital correlation function for the undoped chain
$\langle\vec\tau_i\cdot\vec\tau_{i+1}\rangle^{(0)}=-0.437$ is
modified when doping $x$ increases, and the fraction of bonds with
stronger and weaker fluctuations gradually increases. Using the
stronger and weaker orbital correlations calculated for the $C$-AF
phase both for an undoped chain near the charge defect,
$\langle\vec\tau_i\cdot\vec\tau_{i+1}\rangle_{es(ew)}$
(Sec. \ref{sec:caf}), and for a chain containing a hole,
$\langle\vec\tau_i\cdot\vec\tau_{i+1}\rangle_{hs(hw)}$ (Sec.
\ref{sec:holec}), the average orbital correlations
for strong and weak bonds can be evaluated as follows:
\begin{eqnarray}
\label{tts}
\langle\vec\tau_i\cdot\vec\tau_{i+1}\rangle_s &=&
(1-w_h-w_e)\langle\vec\tau_i\cdot\vec\tau_{i+1}\rangle^{(0)}\nonumber\\
&+& w_e\langle\vec\tau_i\cdot\vec\tau_{i+1}\rangle_{es}
 +  w_h\langle\vec\tau_i\cdot\vec\tau_{i+1}\rangle_{hs}\,,  \\
\label{ttw}
\langle\vec\tau_i\cdot\vec\tau_{i+1}\rangle_w &=&
(1-w_h-w_e)\langle\vec\tau_i\cdot\vec\tau_{i+1}\rangle^{(0)}\nonumber\\
&+& w_e\langle\vec\tau_i\cdot\vec\tau_{i+1}\rangle_{ew}
 +  w_h\langle\vec\tau_i\cdot\vec\tau_{i+1}\rangle_{hw}\,.
\end{eqnarray}
The difference between the stronger and weaker bonds increases with
doping as shown in Fig. \ref{fig:ttdim}(a). At the same time the
average orbital correlation function,
\begin{equation}
\label{tta}
\langle\vec\tau_i\cdot\vec\tau_{i+1}\rangle_a =\frac12\Big(
\langle\vec\tau_i\cdot\vec\tau_{i+1}\rangle_s+
\langle\vec\tau_i\cdot\vec\tau_{i+1}\rangle_w\Big)\,,
\end{equation}
increases somewhat with $x$, indicating overall reduction of orbital
fluctuations with increasing doping.

The FM exchange constants along the $c$ axis also alternate when
the orbital state in dimerized, see Fig. \ref{fig:ttdim}(b). They
follow from the superexchange term in the spin-orbital model Eq.
(\ref{HJ}) by inserting average orbital correlations on strong
(weak) bond into the formula for the exchange constant\cite{Kha04}
which follows from the superexchange along the $c$ axis:
\begin{eqnarray}
\label{jcsa}
J_{cs} \!&=&\! \frac12 J\Big\{
\eta r_1-[r_1-\eta(r_1+r_3)]\nonumber\\
&\times&\!\!
\left(\frac14+\langle\vec\tau_i\cdot\vec\tau_{i+1}\rangle_{s}\right)
-\frac{2}{3}\eta r_3
\langle\vec\tau_i\cdot\vec\tau_{i+1}\rangle_{s}\Big\},\\
\label{jcwa}
J_{cw} \!&=&\! \frac12 J\Big\{
\eta r_1-[r_1-\eta(r_1+r_3)]\nonumber\\
&\times&\!\!
\left(\frac14+\langle\vec\tau_i\cdot\vec\tau_{i+1}\rangle_{w}\right)
-\frac{2}{3}\eta r_3
\langle\vec\tau_i\cdot\vec\tau_{i+1}\rangle_{w}\Big\}.
\end{eqnarray}

Anisotropy between the exchange constants $J_{cs}$ and $J_{cw}$,
shown in Fig. \ref{fig:ttdim}(b), is caused by the alternating
orbital correlations between strong and weak bonds, and increases
with increasing doping $x$. The exchange constants may be thus
expressed by the average exchange $J_a$ and the anisotropy
$\delta_c>0$ as follows:
\begin{eqnarray}
\label{jcs}
J_{cs} &=& J_a(1+\delta_c)\,,\nonumber\\
\label{jcw}
J_{cw} &=& J_a(1-\delta_c)\,.
\end{eqnarray}
The average FM interaction $J_a<0$ is gradually weakened by
reduced orbital fluctuations in a doped system. One finds
$J_a=-0.245J$ and $J_a=-0.229J$ for $x=0$ and $x=0.02$,
respectively. In the present approach the anisotropy increases
linearly with doping $x$ and amounts to $\delta_c=0.117$ at
$x=0.02$. This anisotropy is much weaker than that found in the
undoped YVO$_3$ at $T=85$ K, where $\delta_c=0.35$.\cite{Ulr03}
However, one should keep in mind that the above experimental
result concerns the undoped compound ($x=0$), where the present
mechanism of dimerization is absent and only thermal spin
fluctuations contribute. One expects that the present anisotropy
of the FM exchange constants would be enhanced by the interchain
coupling and by thermal fluctuations, both effects not included in
the present approach, and it would be then higher then in the
undoped compound YVO$_3$.

\subsection{ Transition to dimerized $C$-AF phase }
\label{sec:pht}

The energy of the doped $G$-AF and $C$-AF phase was analyzed in a
similar way --- we evaluated the energy changes
with respect to the undoped phases by weighting the terms arising
for the orbital chains near the charge impurity, either with or
without the hole, and we used thereby the weighting factors $w_h$
and $w_0$, see Eqs. (\ref{weight}). In the undoped YVO$_3$ (at
$x=0$) one finds the $G$-AF phase, but the energy of the $C$-AF is
only larger by a rather small energy $\sim 1$ meV per site, as
estimated before.\cite{Ole07} The
magnetic excitations\cite{Ulr03} and optical
experiments\cite{Kha04} suggest that $J\sim 30$ meV. We therefore
define the energy of the $C$-AF phase at $x=0$ as $E_C(0)=0.03J$
and use it as a parameter below.

\begin{figure}[t!]
\includegraphics[width=7.2cm]{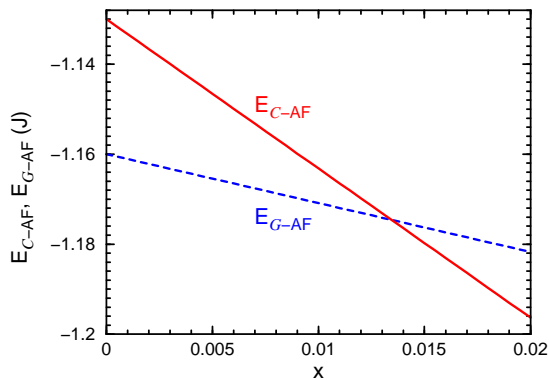}
\caption{(Color online)
Energies of two competing magnetic phases as function of doping
concentration $x$: $G$-AF phase (dashed line) and $C$-AF phase (solid
line) obtained from Eqs.(\ref{egaftot}) and (\ref{ecaftot}), respectively.
Parameters: $V_{ab}=0.2J$, $V_c=0.7J$, $t=6.25J$, $E_C(0)=0.03J$.
}
\label{fig:pht}
\end{figure}

The energies for both phases are given by:
\begin{eqnarray}
\label{egaftot}
E_G(x)&=&w_e E_G^e+w_h E_G^h , \\
\label{ecaftot}
E_C(x)&=&w_0 E_C(0)+w_e E_C^e+w_h E_C^h .
\end{eqnarray}
where $E_G^h\equiv {\cal E}_G(\theta_0)$. The energy increments $E_X^e$
and $E_X^h$, with $X=G,C$, follow from the analysis of the respective
orbital chains without a hole ($E_X^e$) and with a single doped hole
($E_X^h$). The magnetic contributions to the energy
and the energy difference originating from different orbital order in
both undoped phases is already included in $E_C(0)$. The magnetic
order is only influenced locally near the charge defect in the chains
containing a doped hole, and otherwise remains undisturbed, see
Sec. \ref{sec:de}. Therefore we included also the magnetic term in
$E_G^h$ in Eq. (\ref{egaf}).

By evaluating the energy contributions to $E_G(x)$ and $E_C(x)$ we
derived the energies of both phases shown in Fig. \ref{fig:pht}. One
finds that the energy of the $C$-AF phase decreases faster than that of
the $G$-AF phase when doping increases. This results predominantly
from the full contribution of the hopping $t$ on the bond
$\langle 12\rangle$ occupied by the hole in the $C$-AF phase, while
only a fraction of the kinetic energy $t\sin\theta_0$ is released when
it has to compete with robust AF order which hinders it in the $G$-AF
phase.

\section{ Discussion and summary }
\label{sec:summa}

In this paper we addressed the microscopic reasons responsible for
the fast disappearance of the fragile $G$-AF phase with doping in
Y$_{1-x}$Ca$_x$VO$_3$ compounds. Our investigation of the
electronic structure in the $C$-AF phase demonstrated that the
undoped system is a multi-orbital Mott-Hubbard insulator and the
upper Hubbard band consists of three subbands. The multiplet
splitting corresponds to excitations identified in the optical
spectroscopy.\cite{Miy02,Kha04,Kue09} Next we have considered
doped systems and we introduced a model for generic defects in the
perovskite structure. The model is consistent with the
experimentally observed variation of the optical spectra which
show a distinct absorption deep in the Mott-Hubbard
gap.\cite{Fuj08} Our interpretation of this maximum as originating
from the vanadium states localized in the immediate neighborhood
of defects turned out to be consistent with independent estimates
of the binding energy of a hole close the Ca ion. This was further
supported by the spectra obtained by analyzing the occupied and
unoccupied vanadium states within the Hartree-Fock approximation.
In this way we derived large binding energy of the hole being
$\sim 1$ eV, which guarantees that the hole is confined to a cube
built by vanadium ions around the Ca defect.

By reconsidering the local interaction  parameters defining the
multiband Hubbard model for the Y$_{1-x}$Ca$_x$VO$_3$ compounds
and by analyzing the spectral functions obtained in the
Hartree-Fock approach, we arrived at the conclusion that electron
correlations are sufficiently strong to justify the description of
doped materials with the effective spin-orbital $t$-$J$ model.
This model was presented and analyzed for both competing AF
phases: the $G$-AF phase stable at low doping, and the $C$-AF
phase which takes over above the critical doping $x_c\simeq 0.02$.
The crucial observation is that the kinetic energy of the hole is
controlled by the spin configuration, i.e., by a double exchange
mechanism\cite{deG60} similar to that responsible for the
ferromagnetic phase in doped manganites.\cite{Dag01,Feh04,Tok06}
This makes the hopping processes in the two phases distinct along
the $c$ axis, where one phase has FM and the other one AF spin
correlations. In the $ab$ plane both phases have similar AF
correlations, i.e., hole hopping is suppressed here by the spin
order. Thus, the main difference arises from the hopping along the
$c$ axis which is not hindered in the $C$-AF phase and favors this
phase. However, spins in the $G$-AF phase may adjust by tilting
due to the double exchange and allow also for substantial gain of
the kinetic energy. One might expect that the spins would cant to an 
almost FM alignment to favor the kinetic energy as $t\gg J$, where 
$J/4$ is the typical AF superexchange constant between $S=1$ spins, 
which also determines the spin waves in the $G$-AF phase of YVO$_3$.

Yet the double exchange mechanism, i.e., the canting of spins, is not 
controlled by $J$ but by the exchange interactions $I_c$ and $I_{ab}$ 
around the hole site. These interactions couple the $s=1/2$ spin at
a V$^{4+}$ ion with the neighboring $S=1$ spins --- they are much
larger than the energy scale $J$ as these processes arise from
exchange interactions violating Hund's exchange $J_H$ but not 
creating $d^3$ configurations which cost the energy $U$, see Eqs.
(\ref{ex10}). Thus, the canting of spins is opposed by rather strong 
AF interactions and the kinetic energy gain in the $G$-AF phase turns 
out to be only a fraction of the one obtained from the FM bonds in 
the $C$-AF phase.\cite{notede} As a result, the $G$-AF phase is stable 
in a window of low Ca doping, approximately for $0\le x<0.02$. In this 
way we have identified the leading mechanism stabilizing the $C$-AF 
phase in doped Y$_{1-x}$Ca$_x$VO$_3$ which stems from spin-orbital 
physics: free hole hopping $\propto t$ on the bonds in the vicinity of 
the Ca defects accompanied by orbital fluctuations along the remaining 
FM bonds along the $c$ axis.

We have analyzed the differences in spin and orbital correlations
around the defect states between the two AF phases: the $G$-AF and
the $C$-AF phase. In the $G$-AF phase, stable at low doping, no
orbital fluctuations can occur along the $c$ axis due to the
static nature of the $C$-AO order. It is for this reason that the
orbital correlations are rather easily modified in the
neighborhood of the Ca defect and the occupied orbital states
follow the orbital polarization interaction imposed by the defect.
On the contrary, in the $C$-AF phase strong orbital fluctuations
oppose the orbital polarization. Thus the orbitals are prevented
to rotate toward the orientations preferred by the charge-orbital
polarization interactions, as shown in Fig. \ref{fig:tt1}.

As an important byproduct of the present study we have discovered
that dimerization of orbital correlations along the $c$ axis is a
natural consequence of doping and may be considered as {\it
induced by hole defects\/} in the orbital chains. In fact, a doped
hole breaks the orbital chain and generates alternating orbital
bond correlations. Due to the spin-orbital coupling, this
alternation in the orbital subsystem induces the alternation of
stronger and weaker effective FM exchange interactions along the
$c$ axis. Therefore, the performed analysis leads to a prediction
that dimerization of the FM exchange interactions should be {\it
enhanced by doping\/}. We emphasize that this mechanism of
dimerization is distinct from thermal fluctuations that are
necessary to stabilize dimerized spin and orbital interactions in
the undoped spin-orbital chain.\cite{Sir08} We expect that both
mechanisms: (i) orbital-Peierls dimerization,\cite{Sir08} and (ii)
defect induced dimerization of orbital correlations analyzed here,
will support each other in doped Y$_{1-x}$Ca$_x$VO$_3$.

Summarizing, we have shown that a phase transition to the $C$-AF
phase can be explained by the double exchange mechanism using the
spin-orbital physics of the doped $R$VO$_3$ vanadates, as the
kinetic energy of doped holes is much lower in the $C$-AF phase
and this energy gain compensates the loss of the magnetic
superexchange energy already at rather low doping $x\simeq 0.02$.
We have shown that the experimentally observed magnetic
transition\cite{Fuj08} to the $C$-AF phase may be reproduced
within the presented microscopic model using the parameters
consistent with other experimental data.

Another challenging problem, not addressed here, is a surprising
stability of the $C$-AF phase under doping in
La$_{1-x}$Sr$_x$VO$_3$ compounds. The present calculations have
shown that large binding energy confines doped holes to the
immediate neighborhood of Ca ions in Y$_{1-x}$Ca$_x$VO$_3$. If a
similar hole confinement takes place also in La$_{1-x}$Sr$_x$VO$_3$, 
it could explain the insulating state found in these compounds in a 
broad range of doping $0<x<0.18$.\cite{Fuj06} Although the theoretical 
explanation of the evolution of electronic and magnetic properties of
La$_{1-x}$Sr$_x$VO$_3$ has still to be constructed, we note that
large binding energy between holes and defect states is consistent
with these observations. Therefore, we suggest that it is a
necessary ingredient of any realistic theoretical approach to the
doped vanadium perovskites.

\acknowledgments

We thank Bernhard Keimer, Yoshinori Tokura and Clemens Ulrich for 
insightful discussions.  We are grateful to Markus Gr\"uninger and 
Julia K\"uppersbusch for sending us their ellipsometry data prior to 
publication, and to Adolfo Avella for his careful reading of the 
manuscript.
A.M.O. acknowledges support by the Foundation for Polish Science
(FNP) and by the Polish Ministry of Science and Higher Education
under Project N202 069639.

\appendix

\section{Orbital polarization transition}
\label{sec:bond}

In order to get a qualitative insight into the mechanism of local
suppression of orbital fluctuations near the impurity in the
$C$-AF phase we consider a quantum transition to the orbital
polarization on a single bond (between sites $i=1$ and $i=2$ in
Fig. \ref{fig:orbi}) from the singlet state
${\vec\tau}_i\cdot{\vec\tau}_j=-3/4$ to the fully polarized state
with $\langle\tau_i^x\rangle=1/2$. The orbital Hamiltonian
obtained from Eq. (\ref{tJ}) for the present toy orbital model
with a bond $\langle 12\rangle$ along the $c$ axis is
\begin{equation}
\label{bond}
H=Jr_1\left({\vec\tau}_1\cdot{\vec\tau}_2+\frac{1}{4}\right)
-D\left(\tau_1^x+\tau_2^x\right)\,.
\end{equation}
We have chosen here the polarization interaction $-D$ which favors
$\langle\tau_i^x\rangle=1/2$, but an equivalent result is obtained
for $+D$. The Hamiltonian Eq. (\ref{bond}) is easily diagonalized
using the basis $\{|n\rangle\}$:
\begin{eqnarray}
\label{tri1}
|1\rangle&=&a_1^{\dagger}a_2^{\dagger}|0\rangle\,,   \nonumber \\
\label{tri2}
|2\rangle&=&\frac{1}{\sqrt{2}}\left(
a_1^{\dagger}b_2^{\dagger}+b_1^{\dagger}a_2^{\dagger}\right)|0\rangle\,,
 \nonumber\\
\label{tri3}
|3\rangle&=&b_1^{\dagger}b_2^{\dagger}|0\rangle\,,   \nonumber  \\
\label{sin}
|4\rangle&=&\frac{1}{\sqrt{2}}\left(
a_1^{\dagger}b_2^{\dagger}-b_1^{\dagger}a_2^{\dagger}\right)|0\rangle\,,
\end{eqnarray}
where $|0\rangle$ stands for the vacuum. One finds that the
Hamiltonian matrix $H_{nm}\equiv\langle n|H|m\rangle$ takes the
following form,
\begin{equation}
H_{nm}=\textstyle{ \left( \begin{array}{cccc}
\frac{1}{2}Jr_1 & -\frac{1}{\sqrt{2}}\;D & 0 & 0  \\[0.2cm]
-\frac{1}{\sqrt{2}}\;D & \frac{1}{2}Jr_1 & -\frac{1}{\sqrt{2}}\;D & 0 \\[0.2cm]
0 &-\frac{1}{\sqrt{2}}\;D & \frac{1}{2}Jr_1  & 0  \\[0.2cm]
0 & 0 & 0 & -\frac{1}{2}Jr_1
\end{array}
\right),}
\label{h4sz1}
\end{equation}
and the triplet components $\{|1\rangle,|2\rangle,|3\rangle\}$ are
coupled by the orbital polarization term $\propto D$. The eigenvalues
are:
\begin{eqnarray}
\label{lam1}
\lambda_1&=&\frac12 Jr_1\,,          \\
\label{lam23}
\lambda_{2,3}&=&\frac12 Jr_1\pm D\,,  \\
\label{lam4}
\lambda_4&=&-\frac12 Jr_1\,.
\end{eqnarray}
As expected, the splitting between the singlet $|4\rangle$ and the
triplet $\{|1\rangle,|2\rangle,|3\rangle\}$ state is $Jr_1$ at $D=0$.
It decreases linearly with increasing $D$ and one finds a quantum
transition at the critical value of polarization interaction,
\begin{equation}
D_c=Jr_1\,.
\label{Dc}
\end{equation}
This transition is first order and occurs as a level crossing
between the singlet and fully polarized triplet component with
energy $\lambda_{3}=\frac{1}{2}Jr_1-D$. Above the transition (for
$D>Jr_1$) the orbital state is fully polarized by the
charge-orbital interaction and $\langle\tau_i^x\rangle=1/2$ for
$i=1,2$. In this orbital state fluctuations present in the orbital
singlet are suppressed and triplet correlations take over,
$\langle{\vec\tau}_1\cdot{\vec\tau}_2\rangle=1/4$.

The described orbital transition modifies also the magnetic state
in the coupled spin-orbital system. While the fluctuating orbitals
in the singlet state support FM spin correlations, such a
polarized orbital state at $D>D_c$ supports instead AF spin
correlations along the bond in the spin-orbital model Eq. (\ref{HJ}).

\section{ Superexchange for the $d^1-d^2$ bond }
\label{sec:se01}

Here we present the derivation of the superexchange between the
V$^{4+}$ ion generated by a doped hole and its V$^{3+}$ neighbors.
When a hole is doped at a vanadium ion in YVO$_3$, the resulting
spin electronic configuration is $c^1_i$ ($xy^1_i$), corresponding
to $S=1/2$ spin. The superexchange interaction follows from an
interchange of charge between two sites in the excitation process,
$(c^1)_i(c^1(a/b)^1)_j\rightleftharpoons
(c^1(a/b)^1)_i(c^1)_j$, and involves only Hund's exchange $J_H$
due to the intermediate low-spin ($S=0$) excited state
$(c^1(a/b)^1)_i$. Note that these excitations contribute to the AF
superexchange, while the charge transitions between two ions in FM
configuration do not involve any excitation energy and are treated
explicitly by the hopping term Eq. (\ref{Ht}), similar as in
doped manganites.\cite{Ole02}

\begin{figure}[t!]
\includegraphics[width=5.5cm]{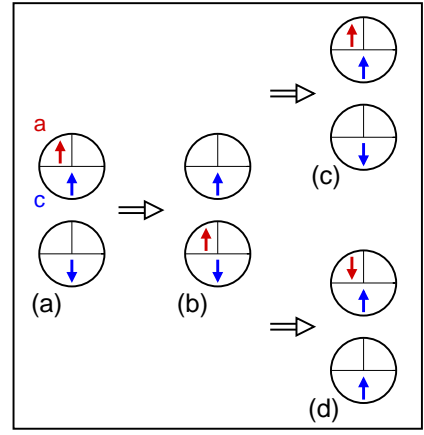}
\caption{(Color online) Artist's view of the virtual charge
excitations $d^1_id^2_j\rightarrow d^2_id^1_j\rightarrow
d^1_id^2_j$ along a bond $\langle ij\rangle\in{\cal C}$ parallel
to the $c$ axis which contribute to the AF superexchange between
V$^{4+}$ and V$^{3+}$ ions in cubic vanadates. After a hopping of
$a$ electron from state (a) to the excited state (b), the
excitation energy $\varepsilon_1=2J_H$ arises. The excitation (b)
may decay in two ways: (c) either the initial configuration is
restored, or (d) spin flips take place at both sites.}
\label{fig:exc}
\end{figure}

The actual occupancy $c^1_i$ and $c^1_j(a/b)^1_j$ of V$^{4+}$ and
V$^{3+}$ ions is responsible for different contributions to the AF
superexchange between the bonds along the $c$ axis and in the $ab$
planes. Consider first a bond $\langle ij\rangle\parallel c$. The
excitations occur here solely by the hopping of an $a$ (or $b$)
electron to the neighboring site occupied by the hole and back, as
shown in Figs. \ref{fig:exc}(a) and \ref{fig:exc}(b). The excited
state, either $c^1_ia^1_i$ or $c^1_ib^1_i$, has to be next
projected on the low-spin ($S=0$) state, with the excitation
energy of $2J_H$. As usual, the final state has the same charge
distribution as the initial one, and the spin configuration is
either the same as in Fig. \ref{fig:exc}(c)], or the $z$-th
components of spins at both sites have been changed by one, see
Fig. \ref{fig:exc}(d). One finds
\begin{eqnarray}
\label{H12c}
H_I^{(c)} &=& \frac{t^2}{4J_H}\sum_{\langle ij\rangle\parallel c}
\left({\vec S}_i\cdot {\vec S}_j-\frac12\right)\nonumber\\
&\times&\left\{
n_i(1-n_j)+n_j(1-n_i)\right\}\,,
\end{eqnarray}
where $n_i$ is the number of electrons in the $\{a,b\}$ orbital
doublet, see Eq. (\ref{ni}).

\begin{figure}[t!]
\includegraphics[width=8cm]{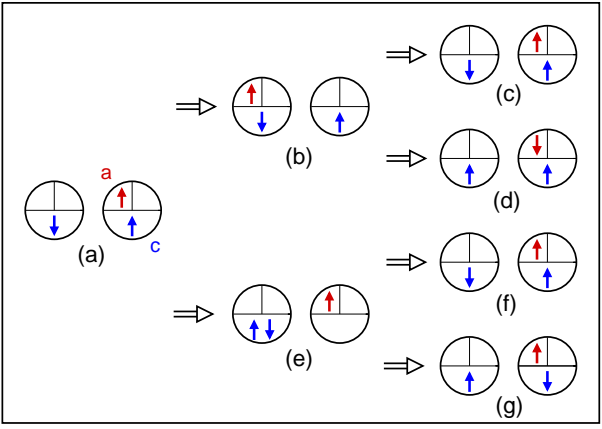}
\caption{(Color online) Artist's view of the virtual charge
excitations $d^1_id^2_j\rightarrow d^2_id^1_j\rightarrow
d^1_id^2_j$ along a bond $\langle ij\rangle\in{\cal C}$ along the
$b$ axis which contribute to the AF superexchange (a) between
V$^{4+}$ and V$^{3+}$ ions in cubic vanadates. When $a$ electron
exchanges with a hole, the excited state (b) with energy
$\varepsilon_1=2J_H$ arises --- it leads to two ground state
configurations, either (c) without or (d) with spin flip. The
hopping of $c$ electron creates a double occupancy (e) with the
excitation energy $\varepsilon_1=5J_H$ --- it gives again two
ground state configurations, either (f) without or (g) with spin
flip. } \label{fig:exa}
\end{figure}

The charge transitions which contribute to the superexchange along
the bonds in $ab$ plane, $\langle ij\rangle\in ab$, have a richer
structure as both $t_{2g}$ electrons at a given V$^{3+}$ ion may
be allowed to hop to the hole site. While the $c$ electron hopping
is allowed along each bond, the second electron is either allowed
to hop or not, depending on its flavor; we show in Fig.
\ref{fig:exa} a case with both electrons contributing to the
superexchange. First, the electron in the degenerate $\{a,b\}$
orbitals may hop to the hole site, see Fig. \ref{fig:exa}(b). This
process is similar to the one for the bond along the $c$ axis, but
is allowed only for half of the bonds, depending on whether the
hopping for the occupied orbital flavor is allowed or not (here we
consider an $a$ electron with the hopping allowed along the $b$
axis). It leads to two final states shown in Figs.
\ref{fig:exa}(c) and \ref{fig:exa}(d). As a new feature one finds
in addition the transitions by both $c$ electrons which create a
double occupancy in $c$ orbital on the undoped site, $c_j^2$, and
leave behind the $a$ (or $b$) electron [Fig. \ref{fig:exa}(e)].
This state has be projected onto the $t_{2g}^2$ eigenstates with
energies $2J_H$ and $5J_H$, see Fig. 1 of Ref. \onlinecite{Ole05}.
The final states, shown in Figs. \ref{fig:exa}(f) and
\ref{fig:exa}(g), have again the same $z$-th spin states as the
initial state, or the spins are flipped. One finds thus the
superexchange,
\begin{eqnarray}
\label{H12a}
H_I^{(ab)} &=&
\frac{t^2}{4J_H}\sum_{\langle ij\rangle\parallel a}
\left({\vec S}_i\cdot {\vec S}_j-\frac12\right)\nonumber\\
&\times&\left\{
n_{ib}(1-n_j)+n_{jb}(1-n_i)\right\}\nonumber\\
&+&\frac{t^2}{4J_H}\sum_{\langle ij\rangle\parallel b}
\left({\vec S}_i\cdot {\vec S}_j-\frac12\right)\nonumber\\
&\times&\left\{
n_{ia}(1-n_j)+n_{ja}(1-n_i)\right\}\nonumber\\
&+&\frac{2t^2}{5J_H}\sum_{\langle ij\rangle\parallel ab}
\left({\vec S}_i\cdot {\vec S}_j-\frac12\right)\nonumber\\
&\times&\left\{n_i(1-n_j)+n_j(1-n_i)\right\}\,.
\end{eqnarray}
The first two terms contribute only when the electron in the
$\{a,b\}$ doublet is allowed to hop along the bond $\langle
ij\rangle$ in the $ab$ plane, while the last term arises from the
$c_i^2$ double occupancies and has no orbital dependence.

Note that after the charge excitation the same orbital
configuration has to be reached in the final state. Therefore,
neither in this case (Fig. \ref{fig:exa}), nor for the bond along
the $c$ axis considered in Fig. \ref{fig:exc}, orbital
fluctuations are present. The orbital dependence occurs in the
interactions derived for the $ab$ plane Eq. (\ref{H12a}), but for
the calculations for 1D orbital chains along the $c$ axis,
performed in this paper, it suffices to average over the orbital
configuration on the sites around the hole. Using the constraint
Eq. (\ref{abc}) one finds:
\begin{eqnarray}
\label{HIc}
H_I^{(c)}\!&=&\! \frac{t^2}{4J_H}\sum_{\langle ij\rangle\parallel c}\!
\left({\vec S}_i\cdot {\vec S}_j-\frac12\right)
\left(2-n_j-n_i\right)\,,\nonumber\\
\label{HIa}
H_I^{(ab)}\! &=&\!
\frac{t^2}{8J_H}\sum_{\langle ij\rangle\parallel ab}\!
\left({\vec S}_i\cdot {\vec S}_j-\frac12\right)\left(2-n_j-n_i\right)
\nonumber\\
\!&+&\!\frac{2t^2}{5J_H}\sum_{\langle ij\rangle\parallel ab}\!
\left({\vec S}_i\cdot {\vec S}_j-\frac12\right)\left(2-n_j-n_i\right)\,.
\end{eqnarray}
This result is used in Sec. \ref{sec:hole} to investigate 1D
orbital chains which contain one doped hole.

\section{ Energy evaluation in the double exchange model }
\label{sec:de12}

Here we present the technical details of the calculations performed in 
Sec. \ref{sec:de}.
The optimal canting angle for the $d^1-d^2$ bond in the $G$-AF phase is
obtained by minimizing the orbital problem Eq. (\ref{gafh}) together 
with the magnetic energy ${\cal E}_G^{\rm mag}(\theta)$ which follows 
from the bonds which are influenced by the spin canting at sites $i=1$
and $i=2$. This energy consists of several contributions due to
the bonds along the $c$ axis and in the $ab$ planes. When
the considered site, either $i=1$ or $i=2$, is undoped, this energy is
determined by the superexchange Eq. (\ref{HJ}) between two $S=1$ spins,
while for the hole site it follows from the superexchange Eq. (\ref{HI})
between the spin $s=1/2$ at the hole site and its $S=1$ neighbor spin.
Using the MF approximation for the superexchange terms one finds:
\begin{eqnarray}
\label{emag}
{\cal E}_G^{\rm mag}(\theta)&=&I_c\sum_{i=1,2}\left\{
\langle s^z_iS^z_{i-1}\rangle+\langle s^z_iS^z_{i+1})\rangle\right\}
(1-n_i)n_j \nonumber\\
&+&I_{ab}\sum_{i=1,2}\sum_{\langle ij\rangle\parallel ab}
\langle s^z_iS^z_j\rangle(1-n_i)n_j\nonumber\\
&+&J_c^s\Big\{(1-n_2)\langle S^z_1 S^z_N\rangle+
(1-n_1)\langle S^z_2  S^z_3\rangle\Big\}
\nonumber\\
&+&J_{ab}^s(1-n_2)\sum_{\langle 1j\rangle\parallel ab}
\langle S^z_1\rangle\langle S^z_j\rangle
\nonumber\\
&+&J_{ab}^s(1-n_1)\sum_{\langle 2j\rangle\parallel ab}
\langle S^z_2\rangle\langle S^z_j\rangle\,.
\end{eqnarray}
The hole disturbs the $G$-AF order locally, so the other bonds are
only weakly influenced and one may evaluate the correlation functions
in Eq. (\ref{emag}) using the classical spin order in this phase, see
Eqs. (\ref{ssg}). When the hole is at site $i=1$, i.e., in the 
$|f\rangle$ state of Fig. \ref{fig:de}(a) (the other $|i\rangle$ 
configuration with a hole at site $i=2$ is equivalent), we have used:
\begin{eqnarray}
\label{s12}
\langle s^z_1S^z_2\rangle+\langle s^z_1S^z_N\rangle&=&
-\frac12\,\cos(2\theta)-\frac12\,\cos\theta\,,
\\
\langle s^z_1S^z_j\rangle&=&-\frac12\,\cos\theta\,.
\\
\langle S^z_2S^z_3\rangle&=&-\cos\theta\,.
\end{eqnarray}
Finally, the ground state of the orbital chain containing one hole
in the $G$-AF phase may be found by minimizing the energy obtained
from the 1D orbital chain Eq. (\ref{gafh}), including the correction 
of the magnetic energy ${\cal E}_G^{\rm mag}(\theta)$ Eq. (\ref{emag}),
\begin{equation}
\label{egaf}
{\cal E}_G(\theta)=\left\langle{\cal H}_{G}^h(\theta)\right\rangle
+{\cal E}_G^{\rm mag}(\theta)\,.
\end{equation}



\begin{thebibliography}{}
\bibitem{notex} These excitations are obtained by diagonalizing the
                on-site Hamiltonian Eq. (\ref{Hee}) in the subspace of
                three degenerate $t_{2g}$ orbitals.
\bibitem{orb2}   We note that the topmost occupied orbital $\alpha=2$ does
                 not change as function of the strength of the orbital
                 polarization field $\Delta_c$.
\bibitem{notede} Note that the exchange bonds around the hole seem to be
                 frustrated in the latter $C$-AF phase, but this spin
                 reorientation does not generate any loss of the
                 superexchange energy since the processes that generate
                 it are absent on the FM bonds, see Eq. (\ref{HI}).
\end{thebibliography}
\end{document}